\documentclass[preprint,12pt]{elsarticle}
\usepackage{graphicx}
\usepackage{amssymb}
\usepackage{amsmath}
\usepackage{subfig,color}
\usepackage{multirow}
\usepackage{colortbl,color}
\usepackage{algorithm}
\usepackage{multirow}
\usepackage{rotating}
\usepackage{amsthm}
\usepackage{hyperref}
\biboptions{comma,sort&compress}
\journal{ }

\begin{document}

\begin{frontmatter}
\title{Phase-coexistence Simulations of Fluid Mixtures by the Markov Chain Monte Carlo Method Using Single-Particle Models}
\author[kaustM+E]{Jun~Li}
\author[kaustNP]{Victor~M.~Calo}
\address[kaustM+E]{Applied Mathematics and Computational Science\\Earth and Environmental Sciences and Engineering\\
King Abdullah University of Science and Technology\\Thuwal, Saudi Arabia}
\address[kaustNP]{Co-Director, Center for Numerical Porous Media\\Applied Mathematics and Computational Science\\Earth and Environmental Sciences and Engineering\\
King Abdullah University of Science and Technology\\Thuwal, Saudi Arabia}
\begin{abstract}
  We present a single-particle Lennard-Jones (L-J) model for~CO$_{2}$ and~N$_{2}$. Simplified L-J models for other small polyatomic molecules can be obtained following the methodology described herein. The phase-coexistence diagrams of single-component systems computed using the proposed single-particle models for CO$_{2}$ and N$_{2}$ agree well with experimental data over a wide range of temperatures. These diagrams are computed using the Markov Chain Monte Carlo~(MC) method based on the Gibbs-$NVT$ ensemble. This good agreement validates the proposed simplified models. That is, with properly selected parameters, the single-particle models have similar accuracy in predicting gas-phase properties as more complex, state-of-the-art molecular models. To further test these single-particle models, three binary mixtures of CH$_{4}$, CO$_{2}$ and N$_{2}$ are studied using a Gibbs-$NPT$ ensemble. These results are compared against experimental data over a wide range of pressures. The single-particle model has similar accuracy in the gas phase as traditional models although its deviation in the liquid phase is greater. The simplified model improves the computational efficiency significantly, particularly in the case of high liquid density where the acceptance rate of the particle-swap trial move increases. The MC method based on Gibbs-$NVT$ ensemble is a viable alternative to simulate phase-coexistence of fluid mixtures. We compare, at constant temperature and pressure, the Gibbs-$NPT$ and Gibbs-$NVT$ ensembles to analyze their performance differences and results consistency. As theoretically predicted, the agreement between the simulations implies that Gibbs-$NVT$ can be used to validate Gibbs-$NPT$ predictions when experimental data is not available.
\end{abstract}
\end{frontmatter}
\section{Introduction}\label{s:intro}

The properties of phase-coexistence are important for many industrial and engineering applications such as the mixture separation through distillation column~\cite{[1]Pucci1986}, the transportation instability due to blockage by natural gas hydrates~\cite{[2]Jr2003} or sulfur deposition~\cite{[3]Al-Awadhy2005}, CO$_{2}$ sequestration~\cite{[4]Herzog2004}, and enhanced oil recovery~\cite{[5]Latil1980}. To obtain these data through experimental observations is time consuming and expensive. Thus, molecular simulations based on the Monte Carlo method are auxiliary tools commonly used to understand phase-coexistence properties.

The Markov Chain Monte Carlo method proposed by Metropolis et al.~\cite{ [6]Metropolis1953} is successful in simulating problems at equilibrium state and here we refer to it as the Monte Carlo (MC) method. It uses the importance sampling idea to generate configurations $\vec X$, which is a high-dimensional vector made up of many molecular positions, according to the probability distribution function $f(\vec X)$. The consecutive configurations constitute a Markov Chain. The MC method estimates the expected values of the quantities of interest by averaging over the sampled configurations. The use of Markov chain makes the algorithm simple and universal but also leads to high correlation of the consecutive samples, which significantly increase the stochastic error in the MC results. Recently, the relationship of the stochastic error with the sample size and sampling interval was analyzed~\cite{[7]Li2012}.

In MC simulations, hundreds or thousands of molecules are distributed inside a cubical box. Periodic boundary conditions are used to analytically enlarge the computational domain as it studies the behavior of a bulk fluid far away from the interface. For problems where the quantities of interest (i.e., pressure, density, mole fraction of each component) depend on molecular position but are independent of the molecular velocity, the MC method records and updates only molecular positions. The MC method based on the Gibbs-$NVT$ ensemble was proposed in~\cite{[8]Panagiotopoulos1987}. It uses two simulation boxes, one for liquid phase and one for the gas phase. The temperature, $T$, the total number of molecules in the two boxes, $N$, and the total volume of the two boxes, $V$, are fixed. The algorithm allows molecules to swap from one phase to the other and volume exchange between the two phases, by changing one box's volume and correspondingly modifying the other's volume keeping the total volume constant. The Gibbs-$NVT$ ensemble MC method effectively simulates phase-coexistence of single component systems but becomes inconvenient in simulating multi-component systems as the pressure is an output of the simulation rather than an input parameter. For the multi-component systems, we use the Gibbs-$NPT$ ensemble MC method~\cite{[9]Panagiotopoulos1988} where the pressure,~$p$, of the two boxes is freely selected and fixed during the simulation. The total volume is not conserved as the volume of each simulation box is changed independently. Many successful applications of the MC method based on Gibbs-$NVT$ and Gibbs-$NPT$ ensembles have been reported in the literature~\cite{[7]Li2012, [10]Errington1998, [11]Smit1995, [12]Martin1998, [13]Nath1998, [14]Errington1999, [15]Potoff1999, [16]Ungerer2006, [17]Hajipour2011, Li2011}.

In this paper, the phase-coexistences of binary mixtures of CH$_{4}$+CO$_{2}$, CH$_{4}$+N$_{2}$ and CO$_{2}$+N$_{2}$ are simulated using a Gibbs-$NPT$ ensemble method. We study the variation with pressure of the mole fraction of each component in the two phases. In order to improve the efficiency of the MC simulation, we neglect the intramolecular structure and model CO$_{2}$ and N$_{2}$ by a single particle as in the traditional model for CH$_{4}$, originally proposed in~\cite{[27]Ravikovitch2001}. The Lennard-Jones parameters for CO$_{2}$ and N$_{2}$ are determined by matching the experimental data in~\cite{[18]Span1996, [19]Span2000} at a temperature far away from the critical temperature and then used in the whole temperature range of interest. The single-particle modeling idea is based on the fact that the reduced equations of state of small molecules are similar to each other. The single-particle model and the selected parameters for CO$_{2}$ and N$_{2}$ are verified first in the simulations of phase-coexistence of single-component systems by comparison with experimental data~\cite{[18]Span1996, [19]Span2000} over a wide range of temperatures. This comparison shows that the single-particle model of CO$_{2}$ with properly selected parameters has similar accuracy in predicting the gas-phase properties as the traditional three-particle model used in~\cite{[20]Harris1995}. To further verify the predictive capabilities of the single-particle model we simulate binary mixtures. As in the single-component case, the MC results using the simplified model agrees well with experimental data~\cite{[22]Kidnay1975, [21]Davalos1976, [23]Somait1978} over a wide range of pressures. We compare the accuracy of the single-particle model against a three-particle model used in~\cite{[24]Do2010} for CO$_{2}$ in the case of the binary mixture of CH$_{4}$+CO$_{2}$. Again, the accuracy of the single-particle model is similar to that of the more complex model in the gas phase. In addition, we present the comparison between the Gibbs-$NPT$ and Gibbs-$NVT$ ensemble MC methods in simulating fluid mixture at the same temperature and pressure. This comparison shows difference in performances between the two algorithms. While comparing the average results of each ensemble method shows that they are consistent with each other under appropriately selected conditions.

\section{Basic algorithm of the Markov Chain Monte Carlo (MC) method}\label{s:MC method}
For problems of equilibrium state, the partition function of the statistical mechanics provides the formula of $f(\vec X)$ and the probability density distribution of the system's configuration $\vec X$ is $f(\vec X)/\int_{\Omega}f(\vec X)d\vec X$. In phase-coexistence problems where the quantities of interest depend only on the molecular position, $\vec X$ is a high-dimensional vector containing the positions of all molecules. The pressure, density, and mole fraction, which depend explicitly on the molecular position, are expressed as the corresponding expected values defined by the following integral:
\begin{equation}\label{eq:<A>=}
    \left<A\right>=\dfrac{\int_{\Omega}f(\vec X)A(\vec X)d\vec X}{\int_{\Omega}f(\vec X)d\vec X}
\end{equation}
where $A(\vec X)$ is the transient value of the quantity of interest at a particular configuration $\vec X$ of the system.  As the formula of $f(\vec X)$ is complicated, it is almost impossible to get an analytical expression for $\left<A\right>$. Traditional quadrature schemes are not applicable due to the large number of nodes required to cover the high dimensional space $\Omega$ where $\vec X$ is defined.

It is convenient to use the Markov Chain Mote Carlo (MC) method~\cite{[6]Metropolis1953} to generate consecutive configurations $\vec X_i$ according to $f(\vec X)$. The MC method uses only $f(\vec X)$ rather than $\int_{\Omega}f(\vec X)d\vec X$. The expected value $\left<A\right>$ is estimated by the average value $\dfrac{1}{n}\sum_{j=1}^nA(\vec X_j)$ over the $n$ sampled configurations $\vec X_j$. The average value converges to the expected value as the sample size $n$ grows infinitely. The algorithm described in~\cite{[25]Frenkel2002} of the Markov Chain Monte Carlo method can be summarized as follows:
\begin{enumerate}
\item Initialization of the configuration $\vec X$: set molecular positions almost uniformly inside the simulation boxes;\label{step1}
\item For each cycle:
  \begin{enumerate}[(a)]
  \item Apply trial move algorithm: the current $\vec X$ is randomly changed to $\vec X'$ by trial moves. The probability density of the event $(\vec X\to\vec X')$ in the trial move is denoted by $\alpha(\vec X\to\vec X')$. To significantly simplify the algorithm the following  symmetric condition \[\alpha(\vec X\to\vec X')=\alpha(\vec X'\to\vec X)\] is required;\label{step2a}
  \item Apply acceptance criterion: the new configuration $\vec X'$ is accepted if $Rf$ (random fraction uniformly distributed in [0, 1]) is less than the acceptance probability $acc(\vec X\to\vec X')$ or rejected otherwise. If rejected, the two consecutive configurations in the Markov Chain are the same. The acceptance probability is equal to \[\min\left[1, \dfrac{\alpha(\vec X'\to\vec X)f(\vec X')}{\alpha(\vec X\to\vec X')f(\vec X)}\right].\] This choice is based on the detailed balance condition for the equilibrium state, namely \[f(\vec X)\alpha(\vec X\to\vec X')acc(\vec X\to\vec X')=f(\vec X')\alpha(\vec X'\to\vec X)acc(\vec X'\to\vec X)\] and the fact that $\dfrac{\min[1, \beta]}{\min[1, \beta^{-1}]}\equiv\beta$. We have that \[acc(\vec X\to\vec X')=\min[1, f(\vec X')/f(\vec X)]\] if the symmetric condition \[\alpha(\vec X\to\vec X')=\alpha(\vec X'\to\vec X)\] holds;\label{step2b}
  \end{enumerate}
\item Sample the system for the quantities $A(\vec X_j)$ of interest after the transitional period required to reach a state of statistical equilibrium is over. Samples are collected every $d$ cycles where $d$ is the sampling interval on the Markov chain. Due to the rejection of trial moves, consecutive samples in the Markov chain are probably identical;
\item Stop once sufficient samples are gathered for analysis. \label{step4}
\end{enumerate}

A detailed analysis leading to choices of $n$ and $d$ that minimize the computational requirements (memory usage and computational time) was presented in~\cite{[7]Li2012}.
\section{MC algorithm based on the Gibbs-$NVT$ and Gibbs-$NPT$ ensembles}\label{s:Gibbs algorithm}
\subsection{Gibbs-$NVT$ ensemble}\label{ss:Gibbs-NVT}

As mentioned above, each molecule is modeled as a single particle and we refer to them simply as particles. For the description with intramolecular structure, the algorithms and formulas described here should be modified accordingly~\cite{[25]Frenkel2002}. A box is employed to represent gas phase and a second one represents the liquid phase while different components can be found in a single box. Two-component systems are discussed here and the notations $a$, $b$ are used to represent different components. The extension to cases with three or more components is straightforward. For the Gibbs-$NVT$ ensemble Monte Carlo method~\cite{[8]Panagiotopoulos1987} introduced in~\cite{[25]Frenkel2002}, we have:
\begin{equation}\label{eq:f=}
\begin{aligned}
    &f\left(\vec S_1,\vec S_2,V_1,V_2,N_{1,a},N_{1,b},N_{2,a},N_{2,b}\right) \\
    &\qquad \qquad\propto\dfrac{V_1^{N_{1,a}+N_{1,b}}V_2^{N_{2,a}+N_{2,b}}\exp\left[-\beta \left(U_1+U_2\right)\right]}{N_{1,a}!N_{1,b}!N_{2,a}!N_{2,b}!}
\end{aligned}
\end{equation}
where $N_{1,a}$ is the particle number of the component $a$ inside the cubic box $1$, $V_1$ is the volume occupied by box $1$, $\vec S_1$ is a high dimensional vector that contains the positions $\vec s_i$ of all particles inside box $1$ normalized by the box size $(V_1)^{1/3}$ (\textit{note}: the subscript $i$ is the particle index and the total particle number inside box $1$ is $N_{1,a}+N_{1,b}$), $\beta=1/(k_\text{B}T)$, $k_\text{B}$ is the Boltzmann constant, $T$ is the temperature, and $U_1=U_1(\vec S_1, V_1)$ is the total potential energy in box $1$ estimated by the summation of pair-wise potential energies $u_{ij}$ contributed by particles $i$ and $j$ contained in the same box. Similar notation applies to the other box and other component in Eq.~\eqref{eq:f=}. The total volume $V_{\text{total}}=V_1+V_2$, total particle numbers $N_{1,a}+N_{2,a}$ and $N_{1,b}+N_{2,b}$ of each component, and temperature $T$ are fixed in the Gibbs-$NVT$ ensemble. The size $L=V^{1/3}$ of the simulation box is very small and the total particle number $N_{1,a}+N_{2,a}+N_{1,b}+N_{2,b}$ is usually only about one thousand due to limitations of computational resources. Thus, periodic boundary conditions are used to analytically enlarge the simulation domain. So, the contribution to the total potential energy $U$ by particle's periodic images is taken into consideration. Taking the box $1$ as an example, the following general form is used to express its energy summation under periodic boundary condition~\cite{[25]Frenkel2002}:
\begin{equation}\label{eq:sumU}
    U_1(\vec S_1,V_1)=\dfrac{1}{2}\ {\sum_{i,j,\vec n}}'u(|\vec r_{ij}+\vec nL_1|)
\end{equation}
where $i$ and $j$ take values from $1$ to $(N_{1,a}+N_{1,b})$ and the factor $1/2$ is used to correct for double counting of the pair-wise contributions, $\vec r_{ij}=L_1\vec s_{ij}=(V_1)^{1/3}(\vec s_i-\vec s_j)$ and $\vec n$ is a vector of three integers from $(-\infty, \infty)$ through which we can represent the contribution by the infinite particle images. For example, if $\vec n=(0,0,1)$, $|\vec r_{ij}+\vec nL_1|=|\vec r_{ij}+(0,0,L_1)|$ which is the distance between particle $i$ and one image of particle $j$. In the simulation, the values of $\vec s_i$ instead of $\vec r_i$ are recorded and so the normalized particle coordinates $\vec s_i$ are unchanged in the trial move of volume change. The prime over the sum notation means that $i=j$ should be excluded when $\vec n=(0,0,0)$, namely we consider the potential energy between particle $i$ and its infinite images but particle $i$ with itself does not contribute to the potential energy. If $i\neq j$, we consider the contribution by particles $i$ and $j$ with $\vec n=(0,0,0)$ as well as the contribution by particle $i$ and the infinite images of particle $j$ with $\vec n\neq(0,0,0)$. Similarly, the transient pressure $p_1$ of box $1$ at a particular configuration $\vec S_1$ is \cite{[25]Frenkel2002}:
\begin{equation}\label{eq:sump}
    p_1=\dfrac{(N_{1,a}+N_{1,b})k_\text{B}T}{V_1}+\dfrac{1}{3V_1}\dfrac{1}{2}{\sum_{i,j,\vec n}}'\left(-\dfrac{du}{dr}r\right)
\end{equation}
where $r=r_{ij}=|\vec r_{ij}+\vec nL_1|$. In the case of Lennard-Jones fluid:
\begin{equation}\label{eq:L-Ju}
    u_{ij}=u_\text{L-J}(r_{ij})=4\epsilon_{ij}\ \left[\left(\dfrac{\sigma_{ij}}{r_{ij}}\right)^{12}-\left(\dfrac{\sigma_{ij}}{r_{ij}}\right)^6\right]
\end{equation}
We specify the values of $\epsilon$ and $\sigma$ for each component of $a$ and $b$. If particle $i$ and $j$ belong to different components, Lorentz-Berthelot's mixing rules are used to compute the cross parameters \[\epsilon_{ab}=(\epsilon_{aa}\epsilon_{bb})^{1/2}\] and \[\sigma_{ab}=\dfrac{(\sigma_{aa}+\sigma_{bb})}2.\]

As introduced in~\cite{[25]Frenkel2002}, a cutoff distance $r_c$, which is smaller than half of the corresponding box size, is employed to simplify the sum operation by limiting the number of terms with $r<r_c$ that need to be calculated explicitly. Here, we use $r_c\equiv0.45L$, which implies that boxes with different sizes have different $r_c$. The value of $r_c$ changes after the accepted trial moves of volume change as $L=V^{1/3}$. So, the contributions by any particle $i$ and its infinite images are neglected as the minimal value of their distances is $L$ and larger than $r_c$. To simplify discussion, we refer to particle $j$ and its infinitely many images as the particle set of $j$. To compute the summation with $i\neq j$ including the potential energy between particle $i$ and the particle set of $j$, we first calculate the normalized distance between particles $i$ and $j$ in each coordinate axis and then get the minimal normalized distance in each coordinate direction while taking the infinite images of particle $j$ into consideration. For example, we first compute the normalized distance $\Delta s_x$ in the $x$ direction by the normalized coordinates $\vec s_i$ and $\vec s_j$. The periodic length of the normalized coordinate at all axes is $1$. Thus, $\Delta s_x-floor(\Delta s_x)$ is positive and belongs to [0, 1), where the function $floor(\Delta s_x)$ returns the maximum integer which is smaller or equal to $\Delta s_x$. Then, the minimal normalized distance $\Delta s_{x,\text{min}}$ in the $x$ direction is $\Delta s_x-floor(\Delta s_x)$ if it is smaller than 0.5 or equal to $1-[\Delta s_x-floor(\Delta s_x)]$ otherwise. The minimal normalized distances $\Delta s_{y,\text{min}}$ and $\Delta s_{z,\text{min}}$ in the $y$ and $z$ directions are computed in the same way. Now, the minimal normalized distance between particle $i$ and the particle set of $j$ is $\Delta s_\text{min}=(\Delta s_{x,\text{min}}^2+\Delta s_{y,\text{min}}^2+\Delta s_{z,\text{min}}^2)^{1/2}$. For any particular $i$ and $j$ ($i\neq j$ as $i=j$ is neglected due to truncation) combined with all possible $\vec n$ in the summation, we only need to check a single pair-wise interaction with the distance equal to $\Delta s_\text{min}$ contribute to the summations in Eqs.~\eqref{eq:sumU}-\eqref{eq:sump} and neglect other infinitely many terms due to the truncation with $r_c<0.5L$. Namely the normalized cutoff distance $s_c$ is smaller than 0.5. According to the above analysis, the number of terms with $r<r_c$ in the summations is finite and its contribution can be computed explicitly.

The neglected contributions with $r>r_c$ to the summations are estimated by tail corrections. These tail corrections for the energy and pressure summations of box 1 are:
\begin{equation}\label{eq:tailU}
    U_1^{\text{tail}}=\sum_{k_1=a}^bN_{1,k_1}\sum_{k_2=a}^b\dfrac{8\pi N_{1,k_2}}{3V_1}\ \epsilon_{k_1k_2}\sigma_{k_1k_2}^3\left[\dfrac{1}{3}\left(\dfrac{\sigma_{k_1k_2}}{r_{c,1}}\right)^9-
   \left (\dfrac{\sigma_{k_1k_2}}{r_{c,1}}\right)^3\right]
\end{equation}
and
\begin{equation}\label{eq:tailp}
    p_1^{\text{tail}}=\sum_{k_1=a}^b\sum_{k_2=a}^b\dfrac{16\pi N_{1,k_1} N_{1,k_2}}{3V_1^2}\epsilon_{k_1k_2}\ \sigma_{k_1k_2}^3 \left[\dfrac{2}{3}\left(\dfrac{\sigma_{k_1k_2}}{r_{c,1}}\right)^9-
      \left(\dfrac{\sigma_{k_1k_2}}{r_{c,1}}\right)^3\right]
\end{equation}
The total energy and pressure are estimated by the sums of the explicit summations with $r<r_c$ and the tail corrections for $r>r_c$. Note that if the components $a$ and $b$ have the same values of $\epsilon$ and $\sigma$, the tail corrections degenerate to
\begin{equation}\label{eq:tailU-single}
    U_1^{\text{tail}}=\dfrac{8\pi (N_{1,a}+N_{1,b})^2}{3V_1}
    \epsilon\ \sigma^3\left[\dfrac{1}{3} \left(\dfrac{\sigma}{r_{c,1}}\right)^9-\left(\dfrac{\sigma}{r_{c,1}}\right)^3\right]  \end{equation}
 and
\begin{equation}\label{eq:tailp-single}
    p_1^{\text{tail}}=\dfrac{16\pi (N_{1,a}+N_{1,b})^2}{3V_1^2}
    \epsilon\ \sigma^3\left[\dfrac{2}{3}\left(\dfrac{\sigma}{r_{c,1}}\right)^9-\left(\dfrac{\sigma}{r_{c,1}}\right)^3\right]
\end{equation}
which are consistent with the results of the single-component system~\cite{[25]Frenkel2002}.

As the configuration $\vec X$ of the probability distribution function of Eq.~\eqref{eq:f=} contains three types of independent variables which are particle coordinates, box volumes, and particle numbers, three kinds of trial moves are necessary: particle displacement, volume change, and particle swap. These satisfy the ergodicity condition which requires that it is possible to visit any $\vec X'\in\Omega$ from the current $\vec X$ in a finite number of trial moves. After getting the total energy $U$ of each box and $f(\vec X)$, the acceptance probability of each trial move can be computed. The three types of trial moves are selected with predetermined probabilities, which can be adjusted during the translational period before reaching the thermal equilibrium state.  The three trial move algorithms used here satisfy the symmetric condition \[\alpha(\vec X\to\vec X')=\alpha(\vec X'\to\vec X)\] and so the acceptance probabilities are determined simply by \[acc(\vec X\to\vec X')=\min\left[1, \dfrac{f(\vec X')}{f(\vec X)}\right].\]

In the trial move of the particle displacement, we select one box denoted by $m$ from boxes $1$ and $2$ with equal probability and then select a particle denoted by $i$ among all particles inside the box $m$ with equal probability. The new normalized coordinate $\vec s'_i$ of the particle $i$ is computed by \[\vec s'_i=\vec s_i+(Rf_1-0.5)\Delta x\] where $Rf_1$ is a random fraction distributed uniformly inside [0, 1] and $\Delta x$ is the step size of the trial move of particle displacement. We denote the new total energy of the box $m$ by $U'_m$. This trial move satisfies the symmetric condition $\alpha(\vec s_i\to\vec s'_i)=\alpha(\vec s'_i\to\vec s_i)$ as $(Rf_1-0.5)$ is distributed uniformly inside [-0.5, 0.5] and so we have:
\begin{equation}\label{eq:acc-1}
    acc(\vec s_i\to\vec s'_i)=\min\{1, \exp[-\beta(U'_m-U_m)]\}
\end{equation}
We change $\vec s_i$ to $\vec s'_i$ if $Rf_2$ is less than $acc(\vec s_i\to\vec s'_i)$ where $Rf_2$ is another uniformly distributed random fraction. After every accepted translational trial moves, the particle $i$ is placed back into the box $m$ by periodic shifting if its new normalized position $\vec s'_i$ is outside the box $m$, namely at least one of its components is outside [0, 1].

In the trial move of volume change, a new variable \[\chi=\ln(V_1/V_2)=\ln[V_1/(V_{\text{total}}-V_1)]\] is introduced~\cite{[25]Frenkel2002} as the total volume $V_{\text{total}}=V_1+V_2$ is constant in the Gibbs-$NVT$ ensemble with  \[f(\vec X)dV_1=f(\vec X)V_1\dfrac{(V_{\text{total}}-V_1)}{V_{\text{total}}}d\chi=g(\vec X)d\chi\] where \(g(\vec X)=f(\vec X)V_1(V_{\text{total}}-V_1)/V_{\text{total}}.\) Thus, we compute a new $\chi'$ by \[\chi'=\chi+(Rf_3-0.5)\Delta V\] where $\Delta V$ is the step size of this trial move. Although the value range of $\chi'$ is $(-\infty, \infty)$, the value of $V'_1$ is always located inside the reasonable range of $(0, V_{\text{total}})$ since \[V'_1=V_{\text{total}}\dfrac{\exp(\chi')}{[1+\exp(\chi')]},\]  and correspondingly, $V'_2=V_{\text{total}}-V'_1$. As the trial move $(\chi\to\chi')$ satisfies the symmetric condition, the acceptance probability of $\chi'$ is computed by $\min[1, g(\vec X')/g(\vec X)]$:
\begin{equation}\label{eq:acc-2}
\begin{aligned}
    &acc(\chi\to\chi')= \\
    &\quad\min\left\{1, \left(\dfrac{V'_1}{V_1}\right)^{N_{1,a}+N_{1,b}+1}\left(\dfrac{V'_2}{V_2}\right)^{N_{2,a}+N_{2,b}+1}
    \exp\left[-\beta\left(U'_1+U'_2-U_1-U_2\right)\right]\right\}
\end{aligned}
\end{equation}
We change $V_1$ and $V_2$ to $V'_1$ and $V_{\text{total}}-V'_1$, respectively, if $Rf_4<acc(\chi\to\chi')$. $\vec S_1$ and $\vec S_2$ are unchanged in this move.

In the trial move of particle swap, we select one box denoted by $m$ from boxes $1$ and $2$ with equal probability to remove a particle $i$ of component $k$. Simultaneously, this particle $i$ is inserted in the other box and placed at a random location. The component $k$ of particle $i$ is selected from components $a$ and $b$ with equal probability here. Generally speaking~\cite{[9]Panagiotopoulos1988}, the component $k$ is selected from all components with predetermined probabilities, which can be adjusted during the transitional period before reaching the thermal equilibrium state. If the particle number $N_{m,k}$ of component $k$ inside box $m$ is zero, the trial move is rejected immediately and the current configuration $\vec X$ is repeated in the Markov Chain. Otherwise, we select a particle denoted by $i$ among those particles of component $k$ inside box $m$ with equal probability. Taking $m=1$ for instance, we remove particle $i$ of component $k$ from box $1$, which changes $N_{1,k}$ to $N_{1,k}-1$ and $U_1$ to $U'_1$. Correspondingly, we create a particle with its coordinate $\vec s_i$ selected randomly and uniformly in box 2, which changes $N_{2,k}$ to $N_{2,k}+1$ and $U_2$ to $U'_2$. This trial move satisfies the symmetric condition and so the acceptance probability is:
\begin{equation}\label{eq:acc-3}
\begin{aligned}
    &acc(N_{1,k}\to N_{1,k}-1) \\
    &\qquad=\min\left\{1, \dfrac{V_2N_{1,k}}{V_1(N_{2,k}+1)}\exp\left[-\beta\left(U'_1+U'_2-U_1-U_2\right)\right]\right\}
\end{aligned}
\end{equation}
The formula for $m=2$ is similar to Eq. \eqref{eq:acc-3}. This trial move is accepted if $Rf_5<acc(N_{1,k}\to N_{1,k}-1)$.

The step sizes $\Delta x$ and $\Delta V$ are adjusted during the transitional period to achieve the prescribed acceptance ratios (0.5 for example) of the corresponding trial moves and fixed later to {\textit{constantly}} satisfy the symmetric condition of trial moves required by the sampling process. The step size of particle swap is fixed at one, namely swapping one particle each time. This makes the acceptance ratio of particle swap fixed and usually very low if the density of the liquid phase is very high. The low acceptance ratio increases the correlation degree of consecutive samples and the statistical variance of the MC results. The single-particle models used here effectively increase the acceptance ratio of particle swap trial move.

The normalized quantities, including the normalized number density $\rho^*=N\sigma^3/V$, volume $V^*=V/\sigma^3$, pressure $p^*=p\sigma^3/\epsilon$, temperature $T^*=Tk_\text{B}/\epsilon$, and energy $u^*=u/\epsilon$, are used in simulations to reduce the numerical error. The normalization parameters $\epsilon$ and $\sigma$ can be freely selected. A convenient selection is to set them equal to the parameters of one component of the mixture.
\subsection{Gibbs-$NPT$ ensemble}\label{ss:Gibbs-NPT}
In the Gibbs-$NPT$ ensemble~\cite{[9]Panagiotopoulos1988}, the total particle numbers $N_{1,a}+N_{2,a}$ and $N_{1,b}+N_{2,b}$ of each component are fixed while the total volume $V_{\text{total}}=V_1+V_2$ is modified during the trial move of volume change. The phase-coexistence pressure $p_{\text{fix}}$ is specified in advance for the two boxes. We denote by $p_{\text{fix}}$ the specified constant used in the following formulas to distinguish it from the value computed by Eq.~\eqref{eq:sump} which is still valid. The computed average pressure by Eq.~\eqref{eq:sump} should converge to the specified value $p_{\text{fix}}$. The original derivation of the Gibbs-$NPT$ ensemble given in~\cite{[9]Panagiotopoulos1988} is based on the precondition of phase-coexistence that the temperature, pressure, and chemical potential of each component are the same for the two phases. As in the Gibbs-$NVT$ ensemble, three kinds of trial moves are used in the Gibbs-$NPT$ ensemble: particle displacement, volume change, and particle swap. The algorithms of particle displacement and particle swap are the same as in the Gibbs-$NVT$ ensemble described above.

For the trial move of volume change, we select one box denoted by $m$ from boxes $1$ and $2$ with equal probability to change its volume while the volume of the other box remains unchanged. Since $V_m$ is the only variable in this trial move, the results of the isothermal-isobaric ensemble is used to obtain the following distribution function:
\begin{equation}\label{eq:f(V)=}
    f(V_m)\propto V_m^{N_{m,a}+N_{m,b}}\exp[-\beta(p_{\text{fix}}V_m+U_m)]
\end{equation}
A new variable $\eta=\ln V_m$ is introduced~\cite{[25]Frenkel2002}. Then, \[f(V_m)dV_m=f(V_m)V_md\eta=q(V_m)d\eta \] where $q(V_m)=f(V_m)V_m.$ We compute a new $\eta'$ as \[\eta'=\eta+(Rf_6-0.5)\Delta V.\] Although the value range of $\eta'$ is $(-\infty, \infty)$, the value of $V'_m$ is always located inside the reasonable range of $(0, \infty)$ as $V'_m=\exp(\eta')$. As the trial move $(\eta\to\eta')$ satisfies the symmetric condition, the acceptance probability of $\eta'$ is computed by $\min[1, q(V'_m)/q(V_m)]$:
\begin{align}\label{eq:acc-4}
    &acc(\eta\to\eta') \nonumber\\
    &\quad=\min\left\{1, \left(\dfrac{V'_m}{V_m}\right)^{N_{m,a}+N_{m,b}+1}\exp\left[-\beta p_{\text{fix}}\left(V'_m-V_m\right) -\beta\left(U'_m-U_m\right)\right]\right\}
\end{align}
We change $V_m$ to $V'_m$ if $Rf_7$ is less than $acc(\eta\to\eta')$.

The two boxes have almost the same uniform initial state. As the volume of each box is changed independently in the Gibbs-$NPT$ ensemble, both boxes are prone to remain in the liquid phase which usually has a lower energy $U$ than the expected gas phase. In this case, it takes a very long computation time for the two boxes to split into different phases which is the final steady state. In order to avoid such sluggish transitional period, we suggest discarding the trial move of volume change during the initial period (for example, the first $10\%$ of the predetermined transitional period) such that the two boxes split quickly into two different phases via the trial move of particle swap. After this initial separation period, three trial moves are selected according to their predetermined probabilities. In addition, the ratio of total particle numbers of the components should be selected such that the mole fraction $(N_{1,a}+N_{2,a})/(N_{1,a}+N_{2,a}+N_{1,b}+N_{2,b})$ is between $x_a$ and $y_a$, which are the steady state mole fractions of component $a$ in the liquid and gas phases, respectively. This requirement also applies to the simulation based on the Gibbs-$NVT$ ensemble for multicomponent systems. For single-component systems, where the Gibbs-$NPT$ ensemble is invalid, the simulation based on the Gibbs-$NVT$ ensemble requires the initial density $\rho$ to lie between the densities of the gas and liquid phases at the equilibrium state.
\section{Parameter determination for the single-particle model}\label{s:L-J parameters}
Usually, CH$_4$ is modeled as a single particle with $\epsilon_{\text{CH}_4}/k_\text{B}=147$ K and $\sigma_{\text{CH}_4}=3.723\times10^{-10}$ m~\cite{[26]Errington1998} while N$_2$ is modeled by two atoms and CO$_2$ by three atoms with fixed bond lengths and bending angle. Following~\cite{[27]Ravikovitch2001}, we model N$_2$ and CO$_2$ by single particles to improve the efficiency of the MC simulation and the parameters of $\epsilon$ and $\sigma$ are selected appropriately to match the existing experimental data. In~\cite{[27]Ravikovitch2001}, the parameters used in the single-particle model were determined according to the mean field approximation to match the experimental data. As will be explained in this section, we advocate a simpler procedure which is easily extensible for other molecules.

The use of single-particle model implies that the equation of state for the normalized quantities \[\rho^*=N\sigma^3/V, \quad p^*=p\sigma^3/\epsilon, \quad T^*=Tk_\text{B}/\epsilon\] is unique for CH$_4$, N$_2$ and CO$_2$ although their parameters of $\epsilon$ and $\sigma$ are different. Using a single-particle model for small molecules like N$_2$ and CO$_2$ is justified by the fact that their reduced quantities $\rho_r=\rho/\rho_c$, $p_r=p/p_c$, $T_r=T/T_c$ roughly satisfy the same reduced equation of state where $\rho_c$, $p_c$, $T_c$ are the critical values of each component. For example, the reduced Peng-Robinson (P-R) equation of state is:
\begin{equation}\label{eq:EOS}
\begin{aligned}
    p_r&=\dfrac{3.2533\rho_rT_r}{1-0.25307\rho_r}-\dfrac{4.839\rho_r^2\alpha(T_r)}
    {1+0.50614\rho_r-0.064044\rho_r^2} \\
    \alpha(T_r)&=\left[1+\left(0.37464+1.54226\omega-0.26992\omega^2\right)\left(1-\sqrt{T_r}\right)\right]^2
\end{aligned}
\end{equation}
where the acentric factor $\omega$ is determined by the critical values~\cite{[28]Firoozabadi1999}. If we neglect the difference due to $\omega$ between different components, the reduced P-R equation of state is unique.

\begin{figure}
  \centering
  \subfloat[Normalized diagram by MC method]{\includegraphics[width=0.45\textwidth]{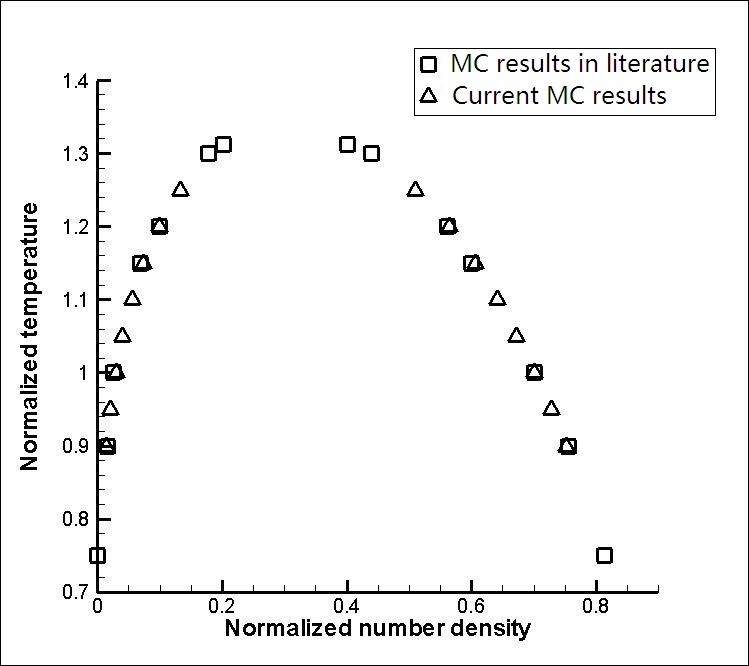}}
  \subfloat[Diagram of methane]{\includegraphics[width=0.45\textwidth]{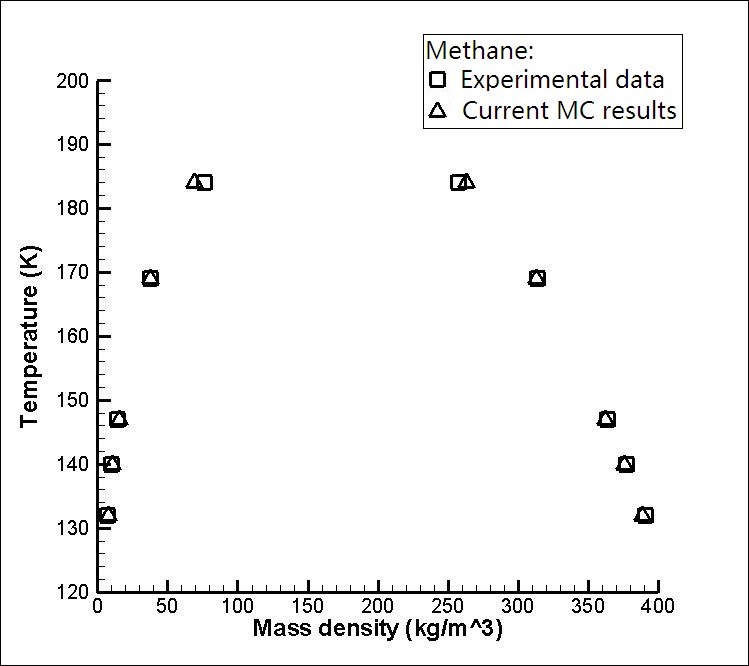}}
  \caption{Transformation of the normalized MC result.}
  \label{fig:TransferData}
\end{figure}

When the molecule is modeled by a single particle, the normalized $\rho^*$, $p^*$, $T^*$ satisfy a unique phase diagram and the related L-J parameters $\epsilon$ and $\sigma$ are used to convert the normalized quantities to values with appropriate physical units. Fig.~\ref{fig:TransferData} left gives the unique phase diagram of the normalized quantities with comparison by the MC results in~\cite{[25]Frenkel2002}. Fig.~\ref{fig:TransferData} right shows the converted results of methane using $\epsilon_{\text{CH}_4}/k_\text{B}=147$ K and $\sigma_{\text{CH}_4}=3.723\times10^{-10}$ m compared by the experimental data~\cite{[29]Setzmann1991}.

\subsection{Parameter selection for CO$_2$}\label{ss:L-J parametersCO2}
For CO$_2$, we select $\epsilon_{\text{CO}_2}$ and $\sigma_{\text{CO}_2}$ appropriately such that the converted results agree with the experimental data~\cite{[18]Span1996}. We choose the normalized numerical results at $T^*=1$ (much lower than the critical $T_c^*\approx1.35$ as in Fig.~\ref{fig:TransferData} left) for converting data since the MC results deviate from experimental data near the critical point. At $T^*=1$, the normalized gas and liquid densities are $\rho_g^*=0.029482$ and $\rho_l^*=0.70111$, respectively, the normalized pressures are $p_g^*=0.024923$ and $p_l^*=0.024896$ (not exactly the same as $p_g^*$ due to stochastic noise). The density ratio is $\rho_l^*/\rho_g^*=23.781$. While, the experimental data~\cite{[18]Span1996} shows that the density ratio $\rho_{l,\text{exp}}/\rho_{g,\text{exp}}$ is $1054.84/43.662=24.159$ at $T=248$ K and $1045.97/46.644=22.425$ at $T=250$ K. We assume that the variation of density ratio with temperature satisfies a linear interpolation and then the density ratio of experimental data at $T=248.4$ K is equal to $23.781$ of the MC simulation at $T^*=1$. This implies that we should select $\epsilon_{\text{CO}_2}/k_\text{B}=T/T^*=248.4$ K such that $T^*=1$ is converted to $T=248.4$ K with the density ratio being closely matched. We use the density of the gas phase to determine another parameter $\sigma_{\text{CO}_2}$ since the stochastic noise in the liquid phase is much larger than that in the gas phase. The experimental mass density of the gas phase is $43.662$ kg/m$^3$ at $T=248$ K and $46.644$ kg/m$^3$ at 250 K. So, the mass density is 44.258 kg/m$^3$ at $T=248.4$ K by interpolation and the corresponding number density is $6.02\times10^{23}\times44.258\times1000/44$ m$^{-3}$ $=6.055\times10^{26}$ m$^{-3}$. As $\rho^*=N\sigma^3/V$ where $N/V$ is the number density, we obtain $\sigma_{\text{CO}_2}=3.652\times10^{-10}$ m which converts the MC result $\rho_g^*=0.029482$ at $T^*=1$ to the experimental data $\rho_{g,\text{exp}}=44.258$ kg/m$^3$ at $T=248.4$ K. Thus, the parameters $\epsilon_{\text{CO}_2}$ and $\sigma_{\text{CO}_2}$ are determined. To further justify this selection, we compute the MC simulation pressure of the gas phase: $p_g=p_g^*\epsilon_{\text{CO}_2}/\sigma_{\text{CO}_2}^3$ $=0.024923\times248.4\times1.380622\times10^{-23}/
(3.652\times10^{-10})^3$ Pa $=1.755\times10^6$ Pa where we used the constant $k_\text{B}=1.380622\times10^{-23}$ J/K. The experimental pressure is $1.6746\times10^6$ Pa at $T=248$ K and $1.785\times10^6$ Pa at 250 K and so it is $1.6967\times10^6$ Pa at $T=248.4$ K by interpolation, which is close to the pressure $1.755\times10^6$ Pa calculated by the MC method with $\epsilon_{\text{CO}_2}/k_\text{B}=248.4$ K and $\sigma_{\text{CO}_2}=3.652\times10^{-10}$ m.

After setting the values of $\epsilon_{\text{CO}_2}$ and $\sigma_{\text{CO}_2}$, we perform MC simulations based on the Gibbs-$NVT$ ensemble at any physical temperature of interest. The MC results at some particular temperatures between 216.592 K of the triple point and 304.1282 K of the critical point are listed in Table~\ref{tab:CO2} and compared with the experimental data~\cite{[18]Span1996} and the MC results in the literature~\cite{[20]Harris1995}, in which CO$_2$ is modeled by three atoms with fixed bond length and each atom has charge. In the elementary physical model (EPM)~\cite{[20]Harris1995}, the bending angle could be fixed or flexible but the results are very close and deviate from the experimental data when $T$ is close to $T_c$. Although the EPM2 obtained by rescaling the parameters of the EPM is proposed in~\cite{[20]Harris1995} to improve the accuracy, the temperature used in the EPM2 is inconsistent with the experimental value. For example, the MC results by the EPM2 at 228 K, 258 K, 298 K agree well with the experimental data at 221 K, 250 K, 289 K, respectively. We choose the MC results by the EPM with fixed bending angle for the comparison of accuracy with the single-particle model used here, since the simulation temperature for EPM can be accurately imposed. Table~\ref{tab:CO2} contains the pressure of gas phase of our simulation and the liquid pressure is neglected due to the large stochastic errors it contains.

\newsavebox{\tableboxa}
\begin{lrbox}{\tableboxa}
\begin{tabular}{ c | c c c | c c c | c c c}
\hline
\multirow {2}{*} {$T$ (K)} &\multicolumn{3}{c|}{Experimental data~\cite{[18]Span1996}} &\multicolumn{3}{c|}{MC results in~\cite{[20]Harris1995}} &\multicolumn{3}{c}{MC results by single-particle model}\\
\cline{2-10}
 & $p$ (MPa) & $\rho_g$ (kg/m$^3$) & $\rho_l$ (kg/m$^3$) & $p$ (MPa) & $\rho_g$ (kg/m$^3$) & $\rho_l$ (kg/m$^3$) & $p$ (MPa) & $\rho_g$ (kg/m$^3$) & $\rho_l$ (kg/m$^3$) \\
\hline
228 & 0.82703 & 21.595 & 1136.34 & 0.76 & 19.3  & 1106  & 0.98264 & 25.4535 & 1116.52 \\
238 & 1.1961  & 31.052 & 1097.05 & 0.95 & 23.7  & 1064  & 1.2938  & 32.8728 & 1086.55 \\
248 & 1.6746  & 43.662 & 1054.84 & 1.49 & 37.5  & 1036  & 1.7573  & 44.3757 & 1054.83 \\
258 & 2.2806  & 60.438 & 1008.71 & 1.98 & 49.4  & 996.9 & 2.2322  & 56.0803 & 1019.97 \\
268 & 3.0334  & 82.965 & 957.04  & 2.62 & 66.8  & 957.3 & 2.8277  & 71.3246 & 983.056 \\
278 & 3.9542  & 114.07 & 897.02  & 3.44 & 89.6  & 909.6 & 3.5732  & 91.4352 & 941.542 \\
288 & 5.0688  & 159.87 & 822.50  & 4.50 & 123.2 & 850.6 & 4.3436  & 112.850 & 896.738 \\
298 & 6.4121  & 240.90 & 712.77  & 5.60 & 164.0 & 776.0 & 5.4119  & 149.593 & 848.864 \\
\hline
\end{tabular}
\end{lrbox}

\begin{table*}[!htb] 
\caption{Comparisons of the phase-coexistence diagrams of CO$_2$ between MC results and experimental data}\label{tab:CO2}
\begin{center}
\resizebox{0.9\textwidth}{!}{\usebox{\tableboxa}}
\end{center}
\end{table*}

\begin{figure}[!htb]
  \centering
  \includegraphics[width=0.45\textwidth]{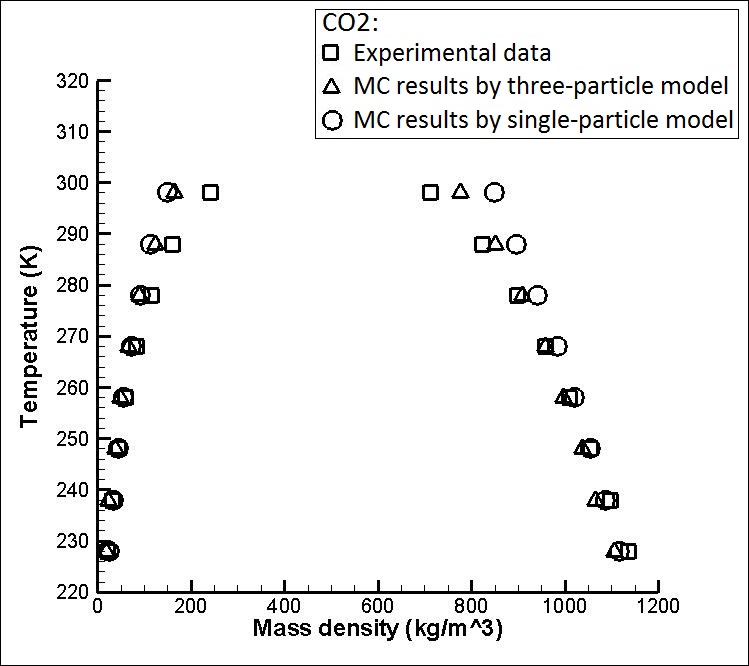}
  \includegraphics[width=0.45\textwidth]{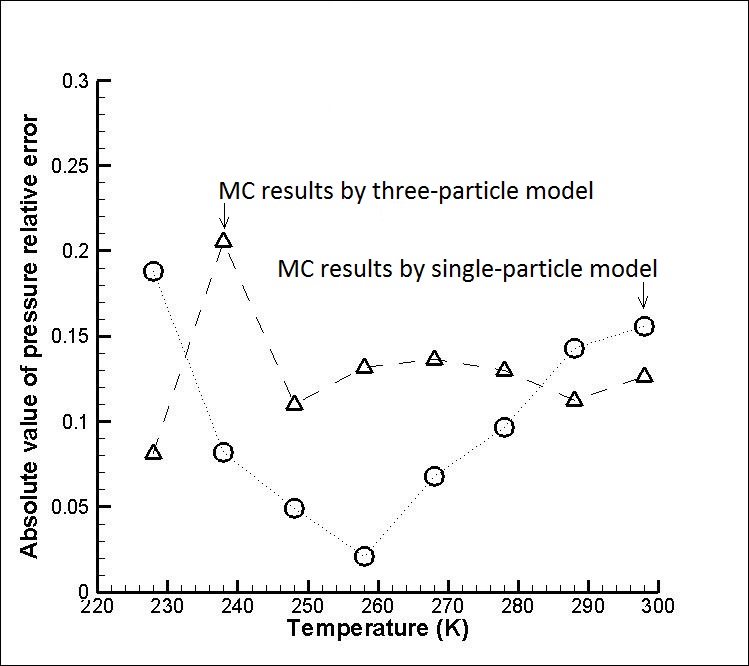} \\
  \includegraphics[width=0.45\textwidth]{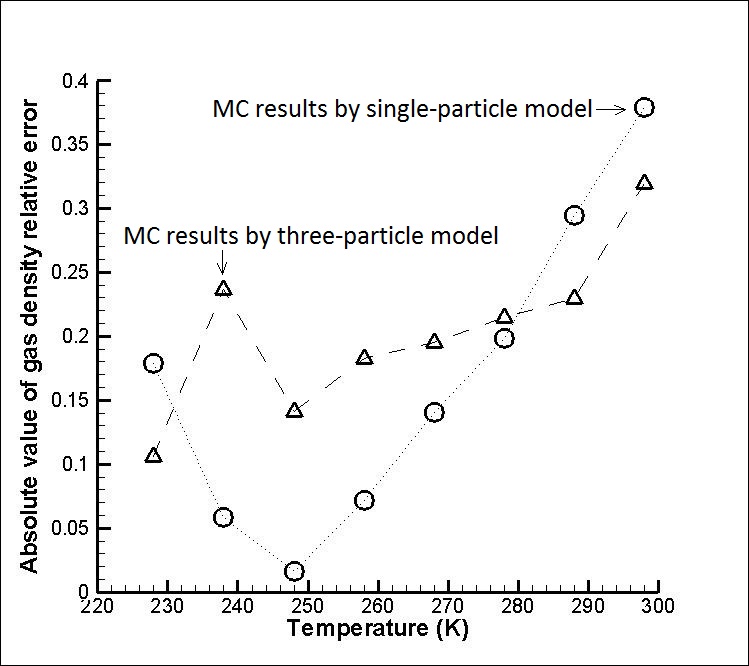}
  \includegraphics[width=0.45\textwidth]{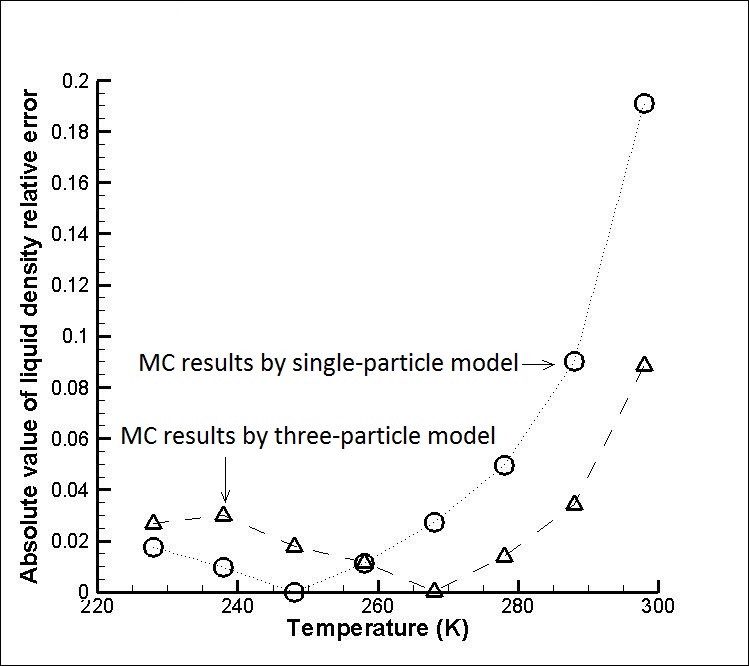}
  \caption{Comparisons of the phase-coexistence diagrams of CO$_2$ and the errors of different molecular models.}
  \label{fig:CO2}
\end{figure}

We use the results in Table \ref{tab:CO2} to compute the relative errors for comparison of accuracy. The relative pressure error is defined as $(p_{\text{sim}}-p_{\text{exp}})/p_{\text{exp}}$ where $p_{\text{sim}}$ and $p_{\text{exp}}$ are the values of MC simulation and experiment, respectively. A similar definition is used for the relative density error. The comparison of the absolute values of the relative errors between the MC results in the literature (three-particle model) and current MC results (single-particle model) are given in Fig. \ref{fig:CO2}.

As shown in Fig. \ref{fig:CO2}, the absolute values of relative errors of the pressure and gas density by the single-particle model used here are smaller than those of the three-particle model in the temperature range from 238 K to 278 K. This is the range where the three-particle model agrees well with the experimental data. The absolute value of relative error of the liquid density by the single-particle model is smaller than that of the three-particle model in the temperature range from 228 K to 258 K. Both models deviate significantly from experimental data when $T$ is close to $T_c\approx304$ K.
\subsection{Parameter selection for N$_2$}\label{ss:L-J parametersN2}
For N$_2$, we select $\epsilon_{\text{N}_2}$ and $\sigma_{\text{N}_2}$ based on the experimental data~\cite{[19]Span2000}. We use the normalized numerical results at $T^*=1$ again and so $\rho_g^*=0.029482$, $\rho_l^*=0.70111$, $\rho_l^*/\rho_g^*=23.781$, $p_g^*=0.024923$. The density ratio $\rho_{l,\text{exp}}/\rho_{g,\text{exp}}$ of the experimental data is 25.276 at $T=98$ K and 23.342 at $T=99$ K. Thus, the experimental density ratio determined using linear interpolation at $T=98.77$ K is equal to 23.781 of the MC simulation at $T^*=1$. This implies that we should select $\epsilon_{\text{N}_2}/k_\text{B}=T/T^*=98.77$ K such that $T^*=1$ is converted to $T=98.77$ K where the density ratio is matched. The experimental density is given in the unit of mol/dm$^3$ and the gas density is 1.0466 mol/dm$^3=29.3048$ kg/m$^{3}$ at $T=98.77$ K by interpolation and so, the corresponding number density is $6.02\times10^{23}\times1.0466\times1000$ m$^{-3}=6.3\times10^{26}$ m$^{-3}$. We select $\sigma_{\text{N}_2}=3.604\times10^{-10}$ m which converts the MC result $\rho_g^*=0.029482$ at $T^*=1$ to the experimental data $\rho_{g,\text{exp}}=29.3048$ kg/m$^3$ at $T=98.77$ K. According to this selection, the gas pressure in MC simulation is \[p_g=p_g^*\epsilon_{\text{N}_2}/\sigma_{\text{N}_2}^3 =0.024923\times98.77\times1.380622\times10^{-23}/(3.604\times10^{-10})^3 \rm{Pa},\] thus,  \(p_g=0.726\times10^6 \rm{Pa}.\) The experimental pressure is $0.67565\times10^6$ Pa at $T=98$ K and $0.72566\times10^6$ Pa at 99 K. Therefore, it is $0.71416\times10^6$ Pa at 98.77 K by interpolation, which is very close to the pressure $0.726\times10^6$ Pa of the MC simulation with $\epsilon_{\text{N}_2}/k_\text{B}=98.77$ K and $\sigma_{\text{N}_2}=3.604\times10^{-10}$ m.

\begin{figure}[!htb]
  \centering
  \includegraphics[width=0.45\textwidth]{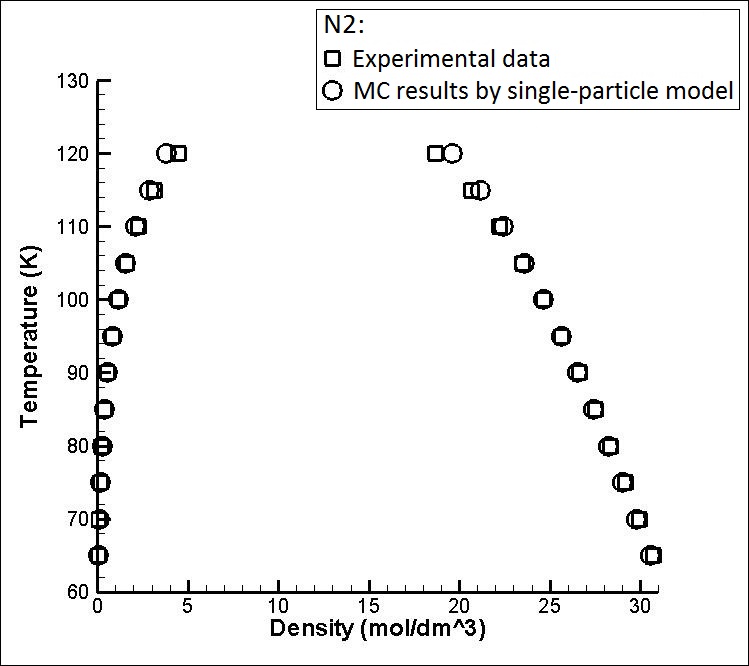}
  \includegraphics[width=0.45\textwidth]{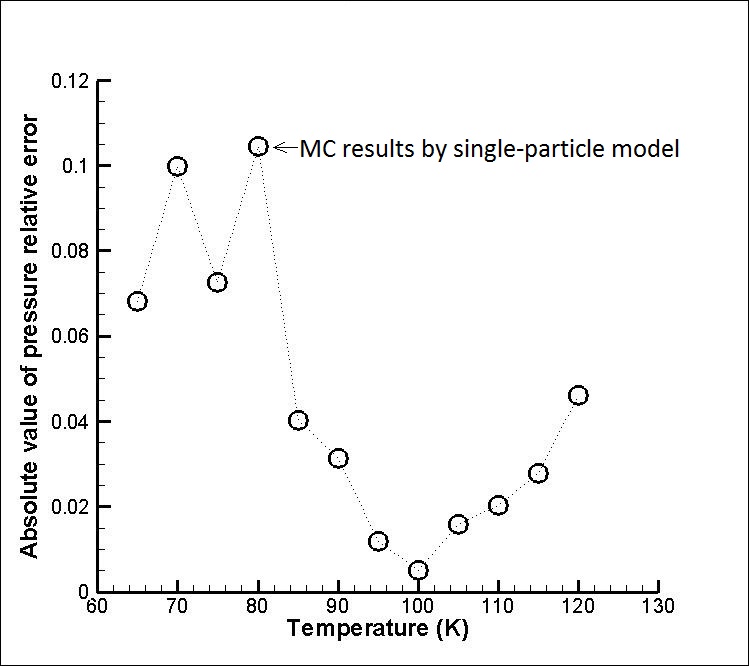} \\
  \includegraphics[width=0.45\textwidth]{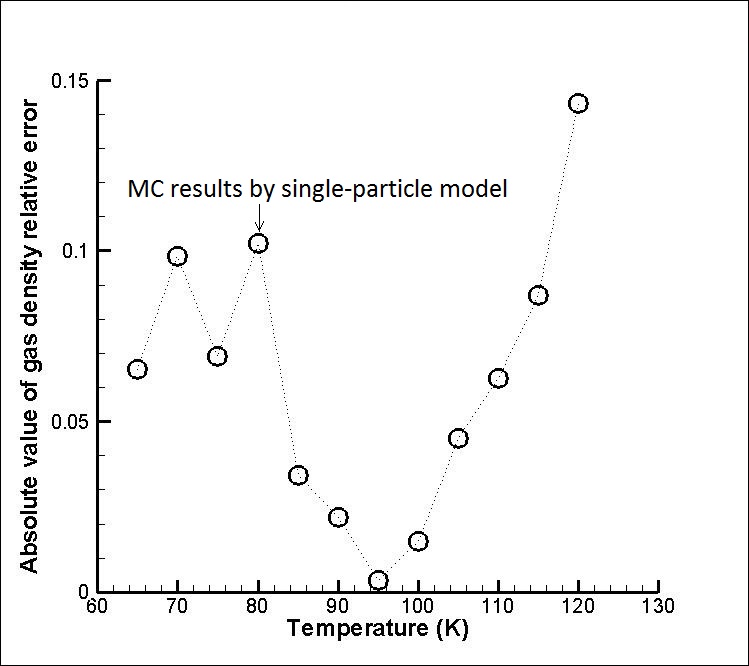}
  \includegraphics[width=0.45\textwidth]{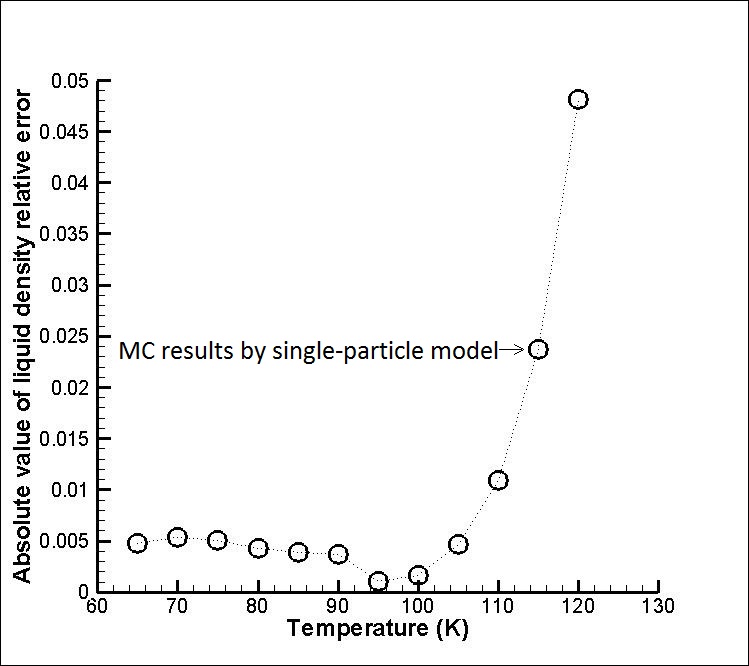}
  \caption{Phase-coexistence diagram of N$_2$ and the error by the single-particle model when compared with the experimental data~\cite{[19]Span2000}.}
  \label{fig:N2}
\end{figure}

We use the values of $\epsilon_{\text{N}_2}$ and $\sigma_{\text{N}_2}$ in the MC simulations at different temperatures between 63.1526 K at the triple point and 126.19 K at the critical point of N$_2$. The comparisons of our MC results by the single-particle model with the experimental data~\cite{[19]Span2000} are listed in Table~\ref{tab:N2} which contains only the pressure of gas phase of our simulations. The corresponding absolute values of the relative errors are plotted in Fig.~\ref{fig:N2}, from which we can see that the agreement of the MC results by the single-particle model with experimental data is better for N$_2$ than CO$_2$.

\newsavebox{\tableboxb}
\begin{lrbox}{\tableboxb}
\begin{tabular}{ c | c c c | c c c }
\hline
\multirow {2}{*} {$T$ (K)} &\multicolumn{3}{c|}{Experimental data~\cite{[19]Span2000}} &\multicolumn{3}{c}{MC results by single-particle model}\\
\cline{2-7}
 & $p$ (MPa) & $\rho_g$ (mol/dm$^3$) & $\rho_l$ (mol/dm$^3$) & $p$ (MPa) & $\rho_g$ (mol/dm$^3$) & $\rho_l$ (mol/dm$^3$) \\
\hline
65  & 0.01740 & 0.03259 & 30.685 & 0.018585 & 0.03472 & 30.537  \\
70  & 0.03854 & 0.06768 & 29.933 & 0.042382 & 0.07434 & 29.772  \\
75  & 0.07604 & 0.12638 & 29.153 & 0.081565 & 0.1351  & 29.003  \\
80  & 0.13687 & 0.21737 & 28.341 & 0.15119  & 0.2396  & 28.218  \\
85  & 0.22886 & 0.35069 & 27.492 & 0.23808  & 0.3627  & 27.384  \\
90  & 0.36046 & 0.53828 & 26.595 & 0.37179  & 0.5500  & 26.497  \\
95  & 0.54052 & 0.79504 & 25.640 & 0.54693  & 0.7923  & 25.611  \\
100 & 0.77827 & 1.1409  & 24.608 & 0.78229  & 1.1238  & 24.648  \\
105 & 1.08331 & 1.6049  & 23.471 & 1.0660   & 1.5324  & 23.581  \\
110 & 1.46581 & 2.2339  & 22.184 & 1.4359   & 2.0940  & 22.425  \\
115 & 1.93704 & 3.1162  & 20.658 & 1.8832   & 2.8450  & 21.148  \\
120 & 2.51058 & 4.4653  & 18.682 & 2.3947   & 3.8254  & 19.580  \\
\hline
\end{tabular}
\end{lrbox}

\begin{table*}[!htb]
\caption{Comparisons of the phase-coexistence diagrams of N$_2$ between MC results and experimental data}\label{tab:N2}
\begin{center}
\resizebox{0.9\textwidth}{!}{\usebox{\tableboxb}}
\end{center}
\end{table*}

Another interesting verification of the parameters selected here is to compare the ratios of the temperatures at the triple and critical points respectively with the ratio of $\epsilon$ which is used for the normalization of temperature. The critical temperature of methane is $T_{c,\text{CH}_4}=190.9$ K and the temperature at its triple point is $T_{t,\text{CH}_4}=90.68$ K. We have \[T_{t,\text{N}_2}/T_{t,\text{CH}_4}=63.1526/90.68=0.696,\] \[T_{c,\text{N}_2}/T_{c,\text{CH}_4}=126.19/190.9=0.661\] and both of them agree well with the ratio of \[\epsilon_{\text{N}_2}/\epsilon_{\text{CH}_4}=98.77/147=0.672.\] For CO$_2$, we have \[T_{c,\text{CO}_2}/T_{c,\text{CH}_4}=304.1282/190.9=1.593\] which agrees well with the ratio of \[\epsilon_{\text{CO}_2}/\epsilon_{\text{CH}_4}=248.4/147=1.69\] but \[T_{t,\text{CO}_2}/T_{t,\text{CH}_4}=216.592/90.68=2.388\] which is due to that the isothermal line at $T=T_t$ of the reduced gas-liquid coexistence area of CO$_2$ is higher than that of CH$_4$.
\section{Simulations of phase-coexistence of binary mixtures}\label{s:Mixture results}
We use the same notation $p$ in the following tables to represent the pressure used in experiments and the parameter $p_{\text{fix}}$ used in MC simulations. The computed pressure by Eq.~\eqref{eq:sump} in MC simulations is given in the next section when comparing the Gibbs-$NVT$ and Gibbs-$NPT$ MC simulations.
\subsection{Gibbs-$NPT$ ensemble MC simulation of CH$_4$+CO$_2$ mixture}\label{ss:CH4+CO2}
\begin{figure}[!htb]
  \centering
  \includegraphics[width=0.45\textwidth]{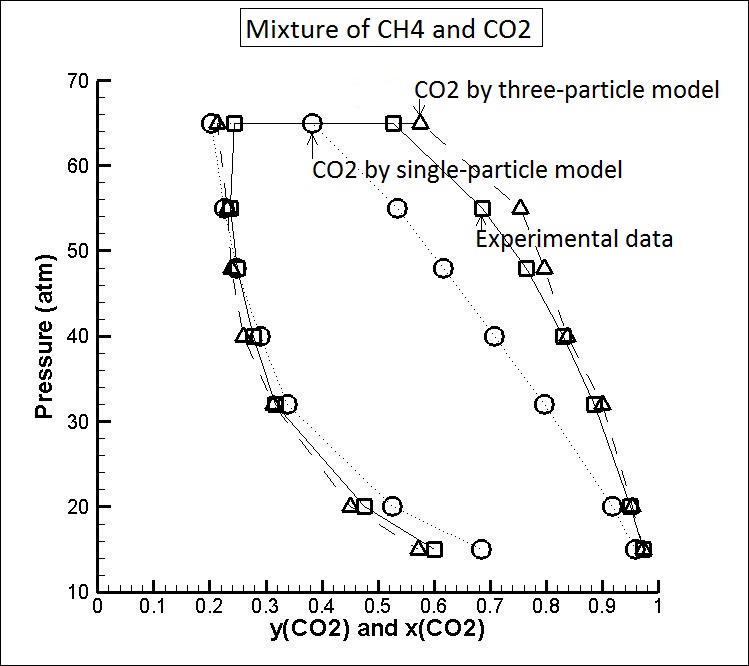}
  \caption{Comparisons of the mole fractions in phase-coexistence of CH$_4$+CO$_2$ at 230 K.}
  \label{fig:CH4+CO2}
\end{figure}

First, we simulate the mixture of CH$_4$+CO$_2$ by the Gibbs-$NPT$ MC method using the single-particle model for CO$_2$. The temperature is fixed at 230 K and the variations of the mole fractions of CO$_2$ in the two phases with the pressure are listed in Table~\ref{tab:CH4CO2} for comparison with experimental data~\cite{[21]Davalos1976} and the MC results using a three-particle model for CO$_2$~\cite{[24]Do2010}. Fig.~\ref{fig:CH4+CO2} plots the data presented in Table~\ref{tab:CH4CO2}. As we can see, the MC results using the single-particle model of CO$_2$ agree well with the experimental data in the gas phase but have larger deviation in the liquid phase (namely in $x_{\text{CO}_2}$) than using a three-particle model. This is consistent with the observation in Fig.~\ref{fig:CO2} where the prediction by the single-particle model of CO$_2$ is worse than the three-particle model EPM~\cite{[20]Harris1995} in liquid phase.

\newsavebox{\tableboxc}
\begin{lrbox}{\tableboxc}
\begin{tabular}{ c | c c | c c | c c}
\hline
\multirow {2}{*} {$p$ (atm)} &\multicolumn{2}{c|}{Experimental data~\cite{[21]Davalos1976}} &\multicolumn{2}{c|}{MC results in~\cite{[24]Do2010}} &\multicolumn{2}{c}{MC results by single-particle model}\\
\cline{2-7}
 & $x_{\text{CO}_2}$ & $y_{\text{CO}_2}$ & $x_{\text{CO}_2}$ & $y_{\text{CO}_2}$ & $x_{\text{CO}_2}$ & $y_{\text{CO}_2}$  \\
\hline
15 & 0.973 & 0.601 & 0.973 & 0.571 & 0.959 & 0.684  \\
20 & 0.950 & 0.475 & 0.954 & 0.451 & 0.917 & 0.526  \\
32 & 0.885 & 0.317 & 0.899 & 0.313 & 0.796 & 0.339  \\
40 & 0.830 & 0.277 & 0.838 & 0.260 & 0.707 & 0.290  \\
48 & 0.765 & 0.249 & 0.796 & 0.238 & 0.616 & 0.247  \\
55 & 0.686 & 0.236 & 0.753 & 0.231 & 0.535 & 0.226  \\
65 & 0.528 & 0.243 & 0.575 & 0.214 & 0.382 & 0.203  \\
\hline
\end{tabular}
\end{lrbox}

\begin{table*}[!htb]
\caption{Comparisons of the mole fractions in phase-coexistence of CH$_4$+CO$_2$ at 230 K}\label{tab:CH4CO2}
\begin{center}
\resizebox{0.9\textwidth}{!}{\usebox{\tableboxc}}
\end{center}
\end{table*}

\subsection{Gibbs-$NPT$ ensemble MC simulation of CH$_4$+N$_2$ mixture}\label{ss:CH4+N2}
\begin{figure}[!htb]
  \centering
  \includegraphics[width=0.45\textwidth]{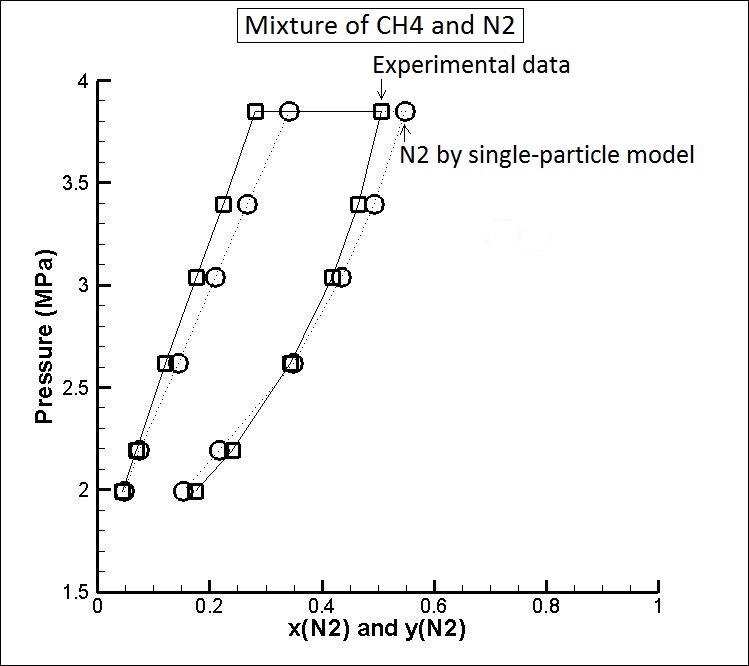}
  \caption{Comparisons of the mole fractions in phase-coexistence of CH$_4$+N$_2$ at 160 K.}
  \label{fig:CH4+N2}
\end{figure}

We also simulate the mixture of CH$_4$+N$_2$ at 160 K and N$_2$ is modeled by a single-particle model. The variation of the mole fraction of N$_2$ in the two phases with the pressure and the corresponding experimental data~\cite{[22]Kidnay1975} are listed in Table~\ref{tab:CH4N2}. The same results are plotted in Fig.~\ref{fig:CH4+N2}. The agreement of our MC results with experimental data in the liquid phase is better than that in the simulation of the mixture of CH$_4$+CO$_2$ by single-particle model for CO$_2$. Nevertheless, the accuracy in the gas phase does not improve in spite of the fact that the MC density results of pure CH$_4$ and N$_2$ agree very well with experimental data (see Figs.~\ref{fig:TransferData} and~\ref{fig:N2}). A possible explanation for this model behavior is that the pressure deviation of the single-particle model for a pure component system increases the computed density (namely mole fraction) deviation as the pressure is an input parameter in the simulation of mixture.

\newsavebox{\tableboxd}
\begin{lrbox}{\tableboxd}
\begin{tabular}{ c | c c | c c }
\hline
\multirow {2}{*} {$p$ (MPa)} &\multicolumn{2}{c|}{Experimental data \cite{[22]Kidnay1975}} &\multicolumn{2}{c}{MC results by single-particle model}\\
\cline{2-5}
 & $x_{\text{N}_2}$ & $y_{\text{N}_2}$ & $x_{\text{N}_2}$ & $y_{\text{N}_2}$  \\
\hline
1.9913 & 0.0448 & 0.1742 & 0.0478 & 0.1521  \\
2.194  & 0.0684 & 0.2406 & 0.0742 & 0.2165  \\
2.619  & 0.1205 & 0.3442 & 0.1438 & 0.3483  \\
3.038  & 0.1756 & 0.4184 & 0.2101 & 0.4340  \\
3.395  & 0.2243 & 0.4657 & 0.2674 & 0.4926  \\
3.846  & 0.2820 & 0.5051 & 0.3414 & 0.5492  \\
\hline
\end{tabular}
\end{lrbox}

\begin{table*}[!htb]
\caption{Comparisons of the mole fractions in phase-coexistence of CH$_4$+N$_2$ at 160 K}\label{tab:CH4N2}
\begin{center}
\resizebox{0.9\textwidth}{!}{\usebox{\tableboxd}}
\end{center}
\end{table*}

\subsection{Gibbs-$NPT$ ensemble MC simulation of CO$_2$+N$_2$ mixture}\label{ss:CO2+N2}
\begin{figure}[!htb]
  \centering
  \includegraphics[width=0.45\textwidth]{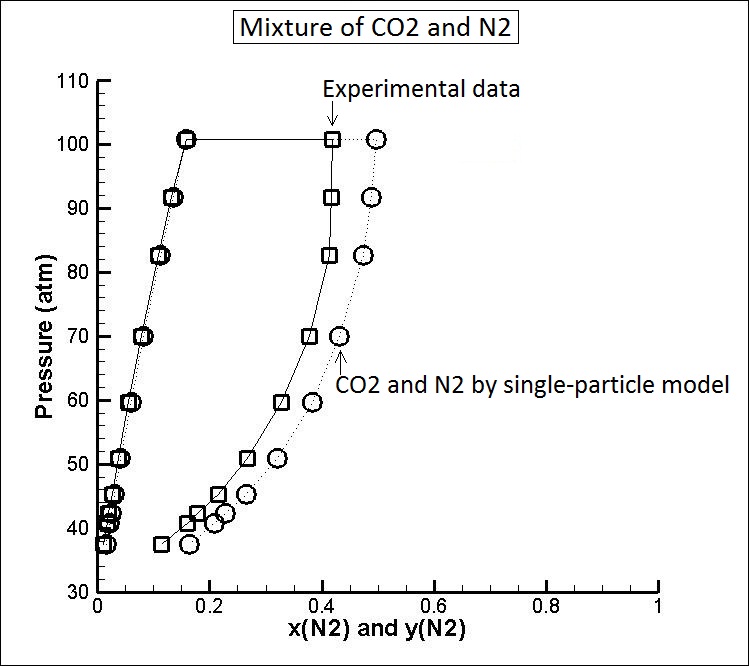}
  \caption{Comparisons of the mole fractions in phase-coexistence of CO$_2$+N$_2$ at 270 K.}
  \label{fig:CO2+N2}
\end{figure}

Finally, we simulate the mixture of CO$_2$+N$_2$ at 270 K. The variations of the mole fraction of N$_2$ in the two phases with pressure and the corresponding experimental data~\cite{[23]Somait1978} are listed  in Table~\ref{tab:CO2N2}. The same results are plotted in Fig.~\ref{fig:CO2+N2}. Although both CO$_2$ and N$_2$ are modeled with the single-particle models, the agreement of our MC results with experimental data in the liquid phase is good, particularly in the case of high pressure. The MC results deviate from the experimental data in the gas phase (namely $y_{\text{N}_2}$) with a shift of about 0.05.

\newsavebox{\tableboxe}
\begin{lrbox}{\tableboxe}
\begin{tabular}{ c | c c | c c }
\hline
\multirow {2}{*} {$p$ (atm)} &\multicolumn{2}{c|}{Experimental data~\cite{[23]Somait1978}} &\multicolumn{2}{c}{MC results by single-particle model}\\
\cline{2-5}
 & $x_{\text{N}_2}$ & $y_{\text{N}_2}$ & $x_{\text{N}_2}$ & $y_{\text{N}_2}$  \\
\hline
37.50  & 0.0108 & 0.1140 & 0.0150 & 0.1637  \\
40.68  & 0.0168 & 0.1598 & 0.0211 & 0.2090  \\
42.25  & 0.0197 & 0.1783 & 0.0238 & 0.2274  \\
45.30  & 0.0263 & 0.2156 & 0.0303 & 0.2660  \\
50.85  & 0.0368 & 0.2674 & 0.0413 & 0.3204  \\
59.70  & 0.0545 & 0.3280 & 0.0596 & 0.3826  \\
70.00  & 0.0778 & 0.3770 & 0.0816 & 0.4317  \\
82.70  & 0.1080 & 0.4126 & 0.1127 & 0.4735  \\
91.70  & 0.1319 & 0.4173 & 0.1343 & 0.4872  \\
100.71 & 0.1585 & 0.4188 & 0.1591 & 0.4976  \\
\hline
\end{tabular}
\end{lrbox}

\begin{table*}[!htb]
\caption{Comparisons of the mole fractions in phase-coexistence of CO$_2$+N$_2$ at 270 K}\label{tab:CO2N2}
\begin{center}
\resizebox{0.9\textwidth}{!}{\usebox{\tableboxe}}
\end{center}
\end{table*}

\section{Comparison between the Gibbs-$NVT$ and Gibbs-$NPT$ ensembles}\label{s:Gibbs-NVT-NPT}
Theoretically, both the Gibbs-$NVT$ and Gibbs-$NPT$ MC methods are valid for simulating mixtures. We simulate the mixture of CH$_4$+N$_2$ at 160 K as an example to show differences between two methods and how their results, under certain conditions, are consistent. We use $\epsilon_{\text{N}_2}$ and $\sigma_{\text{N}_2}$ to normalize quantities including $\rho^*$, $V^*$, $p^*$, $T^*$, $u^*$. To make direct comparison between simulation results, we first run a Gibbs-$NVT$ MC simulation to determine the system's pressure. Using this pressure value as input data, we run a Gibbs-$NPT$ simulation. Similarly to the simulations of the previous section, we choose the total particle number $N$ as 1024 and the normalized total volume $V_{\text{total}}^*$ is 6000 in our Gibbs-$NVT$ MC simulation. $V_{\text{total}}^*$ is selected according to $N$ such that the normalized number density at the initial uniform state lies between the values of the liquid and gas phases in equilibrium state. We set \[N_{1,\text{N}_2}=N_{2,\text{N}_2}=118\] and \[N_{1,\text{CH}_4}=N_{2,\text{CH}_4}=394\] at the initial state making the initial mole fraction of N$_2$ in the two boxes about 0.23 which is between $x_{\text{N}_2}$ and $y_{\text{N}_2}$ at 2.619 MPa. The final pressure obtained by the Gibbs-$NVT$ MC simulation is slightly different from 2.619 MPa because the selections of $V_{\text{total}}^*$, $N_{1,\text{N}_2}$, $N_{2,\text{N}_2}$, $N_{1,\text{CH}_4}$, $N_{2,\text{CH}_4}$ are roughly based on the results of the Gibbs-$NPT$ MC simulation at $p_\text{fix}=2.619$ MPa.

In the Gibbs-$NVT$ MC simulation, the probability for selecting the particle displacement trial move is 0.95, 0.0009 for volume change, and 0.0491 for particle swap after the initial short period of $1\times10^6$ cycles before which the trial move of volume change is avoided. We use $2\times10^8$ cycles for the transitional process and adjust the trial move step sizes $\Delta x$ and $\Delta V$ every $5\times10^5$ cycles during the transitional process such that the acceptance ratios of the trial moves of particle displacement and volume change approach to the predetermined value of 0.5. After the transitional process, the system is sampled every 50 cycles and $2^{24}$ samples are collected to calculate the average values. We get the average values as \begin{align*}
p_g&=2.6793 \rm{MPa}&(p_g^*=0.091976),&\\
p_l&=2.6664 \rm{MPa}& (p_l^*=0.091535),&\\
\rho_{g,\text{N}_2}&=6.1215\times10^{26}\rm{m}^{-3}& (\rho_{g,\text{N}_2}^*=0.028656),&\\ \rho_{l,\text{N}_2}&=1.8245\times10^{27}\rm{m}^{-3}& (\rho_{l,\text{N}_2}^*=0.085406),&\\ \rho_{g,\text{CH}_4}&=1.1004\times10^{27}\rm{m}^{-3}& (\rho_{g,\text{CH}_4}^*=0.051510),&\\ \rho_{l,\text{CH}_4}&=1.0195\times10^{28}\rm{m}^{-3}& (\rho_{l,\text{CH}_4}^*=0.47726).&
\end{align*}

As the MC results in the gas phase contain less stochastic error, we implement the Gibbs-$NPT$ MC simulation at $p_\text{fix}=2.6793$ MPa of the gas pressure of the above Gibbs-$NVT$ MC simulation such that the two simulations are comparable. The parameters setting is almost the same as in the Gibbs-$NVT$ MC simulation but we slightly modified the selection probabilities of the three trial moves to 0.95, 0.0018, 0.0482 having the selection probability of the volume change increased twice because the volume of each box is changed independently in the Gibbs-$NPT$ MC simulation. We get
\begin{align*}
p_g&=2.6796 \rm{MPa}& (p_g^*=0.091986),&\\
p_l&=2.6997 \rm{MPa}& (p_l^*=0.092678),&\\
 \rho_{g,\text{N}_2}&=6.1528\times10^{26} \rm{m}^{-3}& (\rho_{g,\text{N}_2}^*=0.028802),&\\ \rho_{l,\text{N}_2}&=1.8237\times10^{27} \rm{m}^{-3} &(\rho_{l,\text{N}_2}^*=0.085373),&\\ \rho_{g,\text{CH}_4}&=1.0978\times10^{27} \rm{m}^{-3} &(\rho_{g,\text{CH}_4}^*=0.051390),&\\ \rho_{l,\text{CH}_4}&=1.0204\times10^{28} \rm{m}^{-3} &(\rho_{l,\text{CH}_4}^*=0.47768), &
\end{align*}
which are very close to the results of the above Gibbs-$NVT$ MC simulation. The computed gas pressure $p_g=2.6796$ MPa agrees very well with the prescribed parameter $p_\text{fix}=2.6793$ MPa.

\begin{figure}
  \centering
  \includegraphics[width=0.37\textwidth]{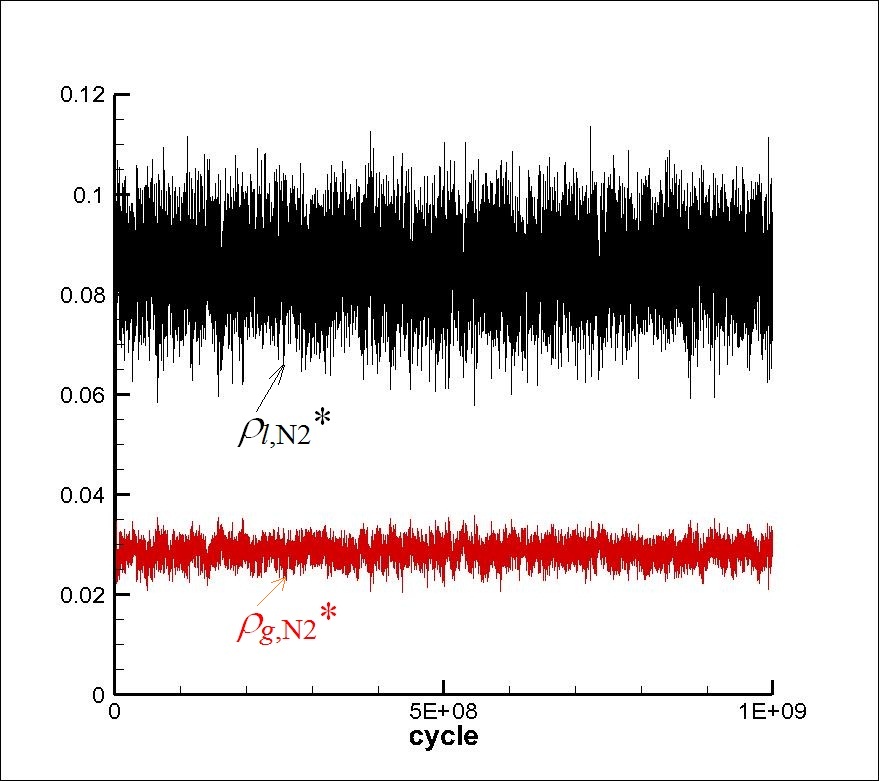}
  \includegraphics[width=0.37\textwidth]{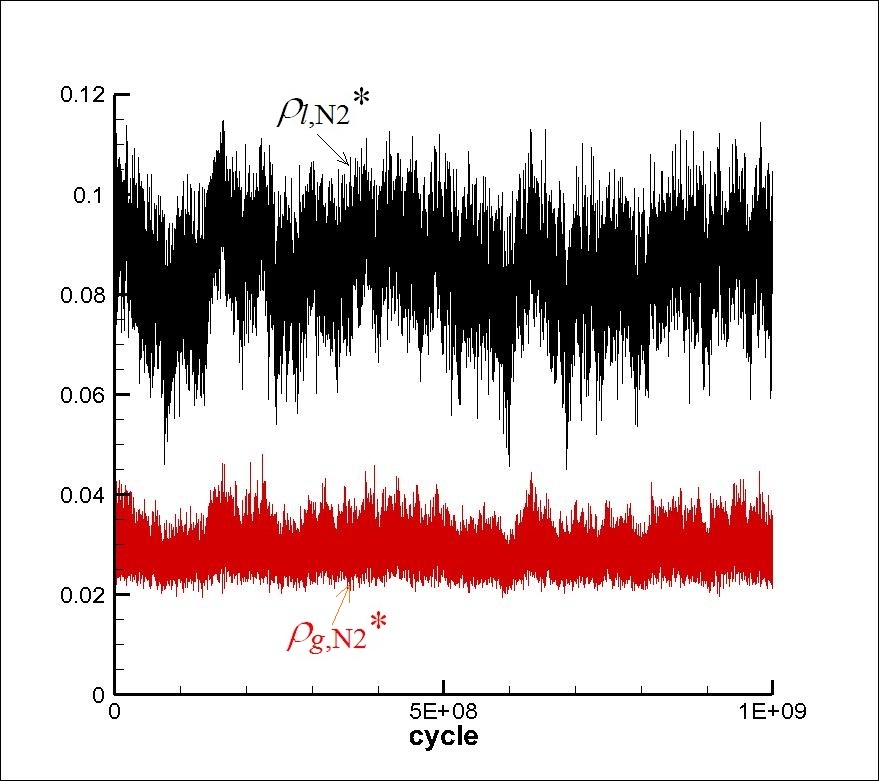} \\
  \includegraphics[width=0.37\textwidth]{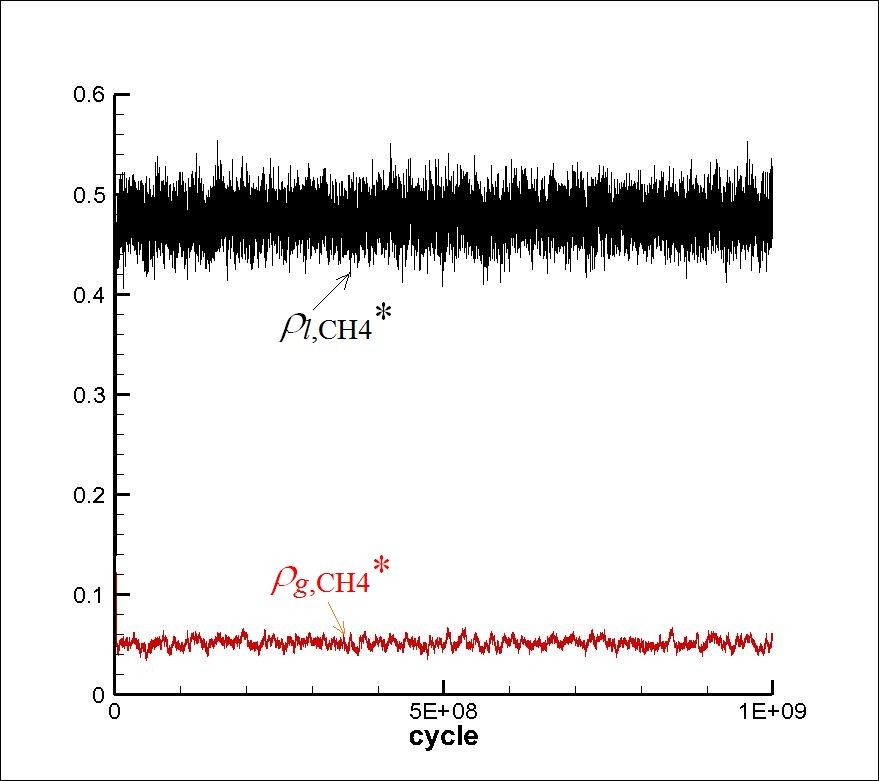}
  \includegraphics[width=0.37\textwidth]{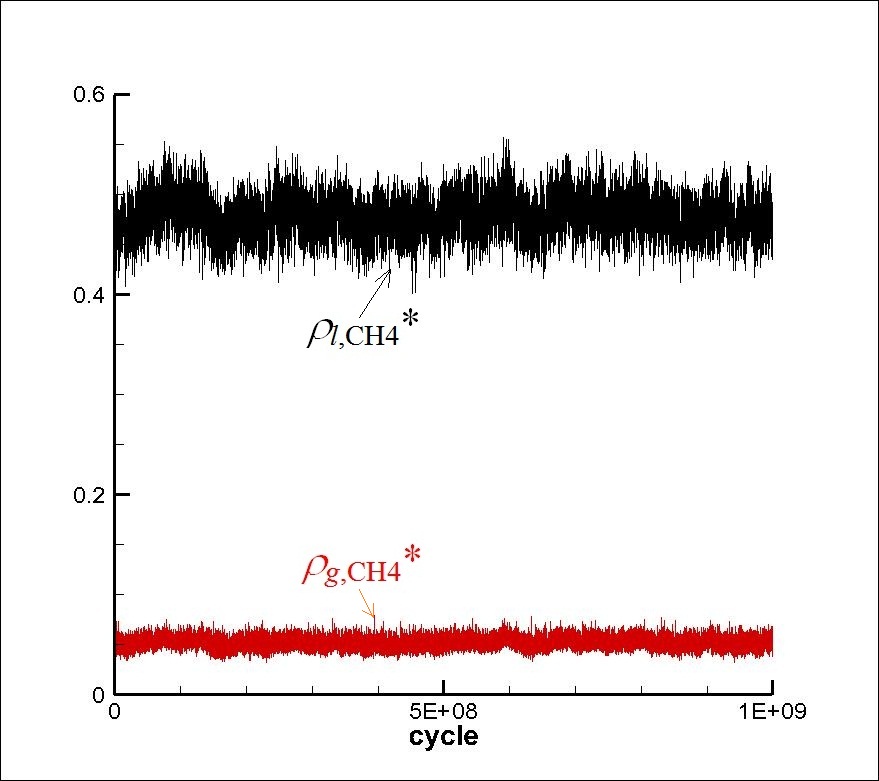} \\
  \includegraphics[width=0.37\textwidth]{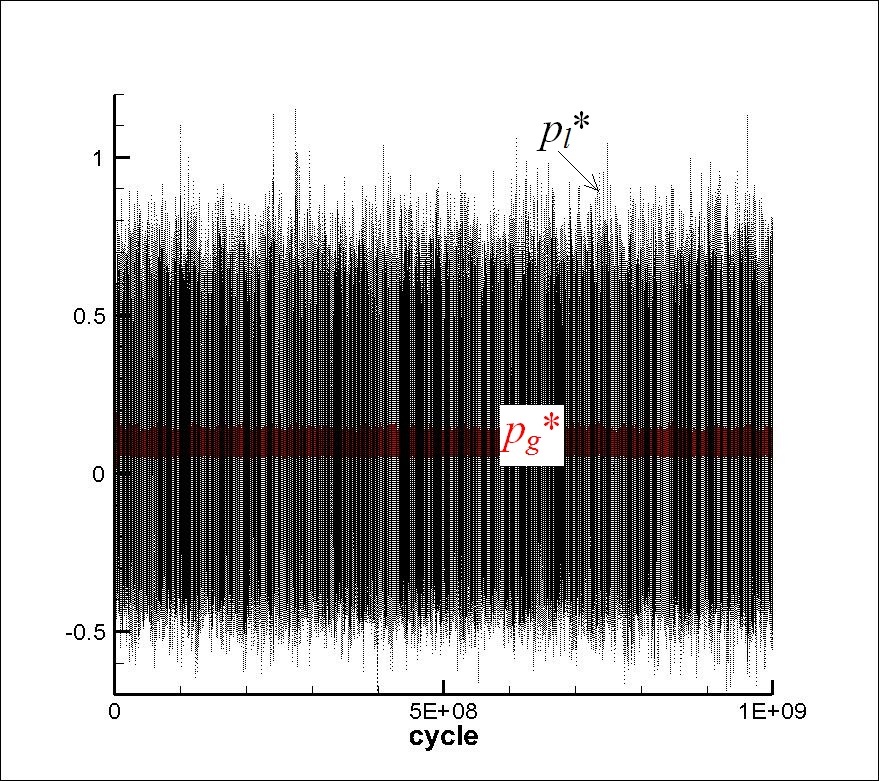}
  \includegraphics[width=0.37\textwidth]{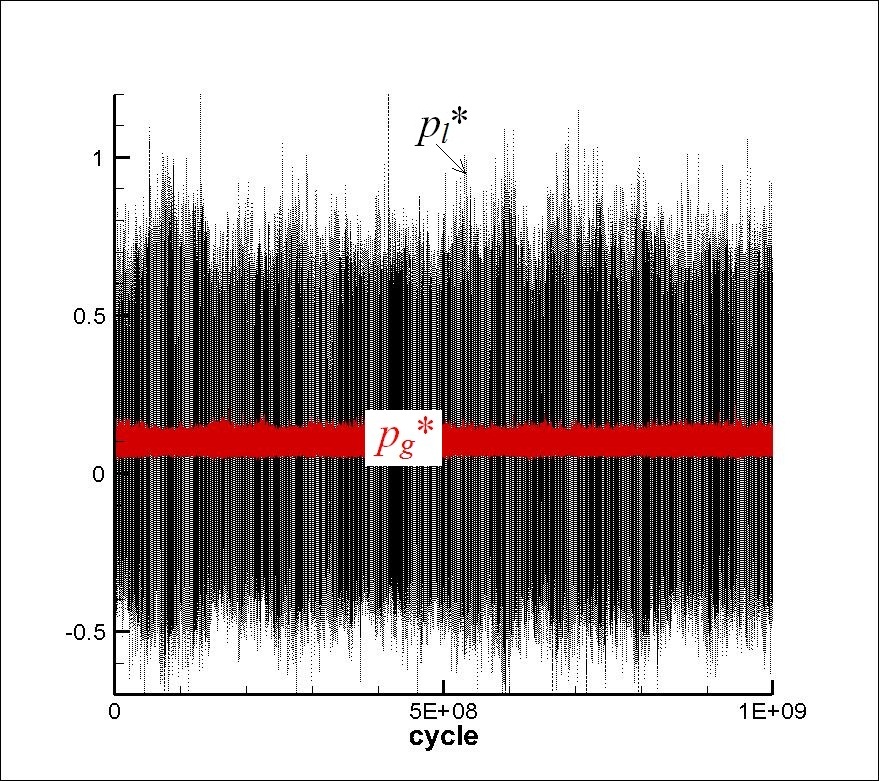} \\
  \includegraphics[width=0.37\textwidth]{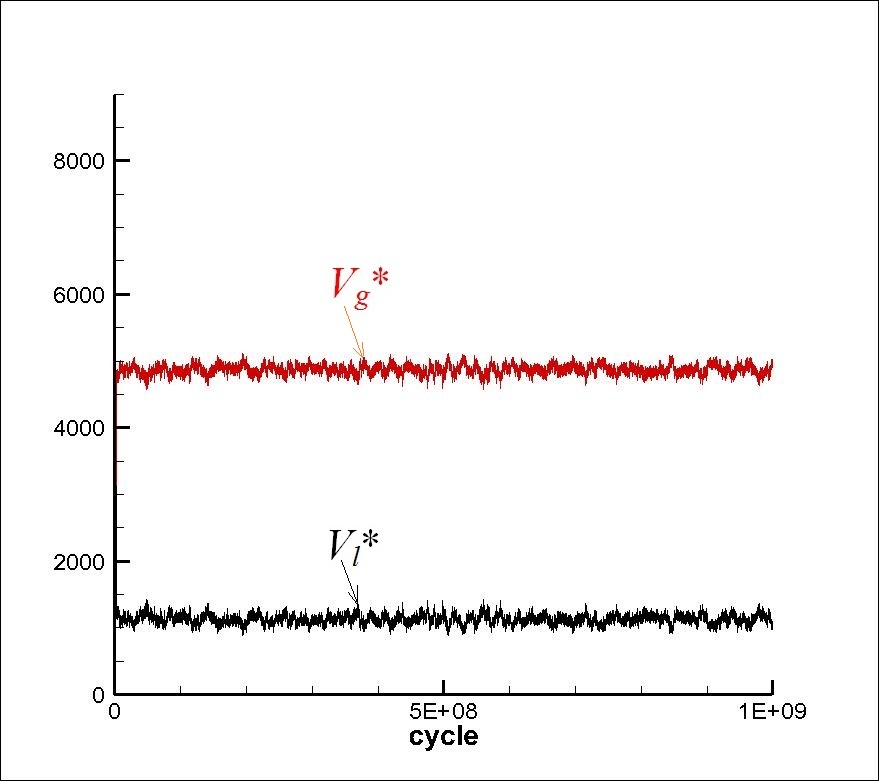}
  \includegraphics[width=0.37\textwidth]{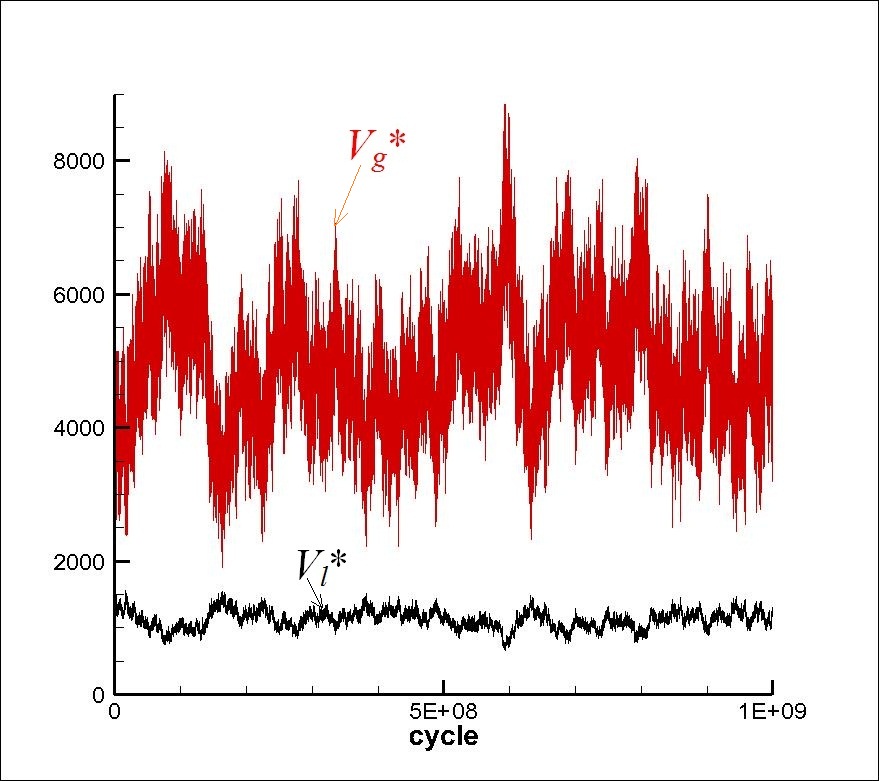} \\
  \caption{Comparison between the Gibbs-$NVT$ (left) and Gibbs-$NPT$ (right) MC simulations of CH$_4$+N$_2$ at 160 K and 2.6793 MPa.}
  \label{fig:Gibbs-NVT-NPT}
\end{figure}

The evolution of the normalized $\rho_{g,\text{N}_2}^*$, $\rho_{l,\text{N}_2}^*$, $\rho_{g,\text{CH}_4}^*$, $\rho_{l,\text{CH}_4}^*$, $p_g^*$, $p_l^*$, $V_g^*$ and $V_l^*$ are given in Fig. \ref{fig:Gibbs-NVT-NPT} to show the comparison between the Gibbs-$NVT$ and Gibbs-$NPT$ MC simulations. The average values agree well with each other but the transient results of the Gibbs-$NPT$ MC simulation contain larger stochastic error particularly in the transient volumes of the two boxes as they are changed independently. But, the application of the Gibbs-$NVT$ MC method in the simulation of mixture is inconvenient because the pressure cannot be prescribed before the simulation is performed and so the study of the relationship between the mole fraction and the pressure at a fixed temperature is inconvenient. Nevertheless, the Gibbs-$NVT$ MC simulation can be used for the validation of Gibbs-$NPT$ MC simulation when experimental data is not available since their simulation results should be consistent with each other.

\section{Conclusions}\label{s:Conclusion}
Markov chain Monte Carlo (MC) simulations of N$_2$ and CO$_2$ are performed using a single-particle model to improve the efficiency of the MC simulation. The corresponding Lennard-Jones~(L-J) parameters are determined according to existing experimental data. The L-J parameters for other small molecules can be obtained using the procedure described herein.

The validity of the single-particle model with the selected parameters is verified in the simulations of systems of pure components and fluid mixtures by comparison with experimental data. In the pure system of N$_2$, the pressure and the gas and liquid densities by the single-particle model agree very well with the experimental data over a wide range of temperatures. For CO$_2$, the single-particle model has comparable accuracy to  the traditional three-particle model in predicting the gas-phase properties but has larger deviation in the liquid phase. In the simulations of binary mixtures of CH$_4$+CO$_2$ and CH$_4$+N$_2$, the predictions by the single-particle model are relevant for  the gas phase although the deviation is obvious in the liquid phase. Nevertheless, the prediction by the single-particle model in the liquid phase becomes better than that in the gas phase when simulating the mixture of CO$_2$+N$_2$.

The comparison between the Gibbs-$NVT$ and Gibbs-$NPT$ MC simulations is made in a particular case of binary mixture to show their difference in performance as well as the consistency of the average results at appropriate conditions. Although the application of the Gibbs-$NVT$ MC simulation is inconvenient for systems of mixtures, it is a useful tool for checking the validity of Gibbs-$NPT$ MC simulation when experimental data is not available.

\end{document}